\documentclass[11pt,a4paper,superscriptaddress]{article}
\pdfoutput=1
\usepackage{jheppub_kim}
\usepackage{rotating} 
\usepackage{graphicx,epsfig}
\usepackage{amsmath}
\usepackage {amssymb}
\usepackage{subfigure}
\usepackage{color}

\usepackage{relsize}

\def\beq{\begin{equation}}
\def\eeq{\end{equation}}
\def\bea{\begin{eqnarray}}
\def\eea{\end{eqnarray}}
\def\be{\begin{equation}}
\def\ee{\end{equation}}
\def\pr{\partial}
\def\nno{\nonumber}
\def\bse{\begin{subequations}}
\def\ese{\end{subequations}}

\graphicspath{{./figs/}}

\begin{document}

\title{Studying G-axion Inflation model in light of PLANCK }

\author{Debaprasad Maity and}

\author{Pankaj Saha}
 
 \affiliation{Indian Institute of Technology Guwahati, Guwahati 781039, India}
 

\emailAdd{debu@iitg.ac.in}

\emailAdd{pankaj.saha@iitg.ac.in}

\abstract{ With the Planck 2015 result, most of the well known canonical large field inflation models turned out to be strongly disfavored. Axion inflation is one of such models which is becoming marginalized with the increasing precession of CMB data.    
	In this paper, we have shown that with a simple Galileon type modification to the marginally favored axion model calling G-axion, we can turn them into one of the most favored models with its detectable prediction of $r$ and $n_s$ within its PLANCK $1\sigma $ range for a wide range of parameters. Interestingly it is this modification which plays the important role in turning the inflationary predictions to be independent of the explicit value of axion decay constant $f$. However, dynamics after the inflation turned out to have a non-trivial dependence on $f$. For each G-axion model there exists a critical value of $f_c$ such that for $f>f_c$ we have the oscillating phase after inflation and for $f<f_c$ we have non-oscillatory phase. Therefore, we obtained a range of sub-Planckian value of model parameters which give rise to consistent inflation. However for sub-Planckian axion decay constant the inflaton field configuration appeared to be singular after the end of inflation. To reheat the universe we, therefore, employ the instant preheating mechanism at the instant of first zero crossing of the inflaton. To our surprise, the instant preheating mechanism turned out to be inefficient as opposed to usual non-oscillatory quintessence model. For another class of G-axion model with super-Planckian axion decay constant, we performed in detail the reheating constraints analysis considering the latest PLANCK result.   
}

\keywords{Galileon Cosmology, Natural Inflation, Instant Preheating}

\maketitle

\newpage
\section{Introduction}\label{intro}
The Inflationary paradigm has been an essential part of standard model cosmology since its inception  
\cite{guth,linde1,steinhardt}. With increasingly precise measurements of various cosmological observables, the inflationary mechanism as the initial condition of standard $\Lambda$CDM cosmology is gradually becoming unique in nature. However, it is the source of inflation, which remains obscured till today because of nearly universal tree level prediction of the inflationary observables for a large number of models, \cite{Martin:2013tda}.
According to the latest cosmological measurements made by Planck\cite{PLANCK}, Keck Array, and BICEP2  
Collaborations\cite{Array:2015xqh}, the scalar spectral index of curvature perturbation is 
$n_s = 0.968 \pm 0.006$ and the scale dependence of scalar spectral index is tightly constrained 
to $d n_s/d \ln k =-0.003 \pm 0.007$. The upper bound on the tensor-to-scalar ratio, 
for 95 GHz Data From Keck Array, is $ r_{0.05} < 0.07 \text{(95 \% CL)}$. All these cosmological 
observables can be obtained by introducing the inflation mechanism driven by either single or multiple scalar fields known as inflaton with a large variety of potential. It is based on the idea of slow-roll dynamics where the inflation rolls down an almost flat potential for long time to solve the so-called 
horizon and homogeneity problem(see\cite{baumann-tasi} for a review). In terms of classical model building, it is very simple to construct such a flat potential. However, such an effective flat potential is very difficult to achieve in the framework of quantum field theory due to \textit{naturalness}. 
Therefore, it is instructive to invoke shift symmetry in the inflaton field space.
Importantly the constant shift symmetry naturally provides nearly flat potential and makes it stable against radiative correction. This was the idea behind the so-called natural inflation model\cite{freese} introduced in the early 90s. 
The inflaton, in this case, is known as axion whose nearly flat potential is of the form  

$$V(\phi) = \Lambda^4 \left[1 - \cos\left(\frac{\phi}{f}\right) \right].$$
Where $\Lambda$ and $f$ are two mass scales characterizing the height($2\Lambda^4$) 
and width($\pi f$) of the potential. The scale $f$ is known as axion decay constant. 
Though it is one of the best theoretically motivated model of inflation, 
it turned out to be observationally tightly constrained. Furthermore, it's predictions closest 
to the observations become quantum field theoretically implausible for the super-planckian 
axion decay constant $f$. Nonetheless, because of its naturalness, many different modifications have been proposed to make it observationally favored and simultaneously bring down the value of $f$ to a sub-Planckian value. String inspired $\mathcal{N}$-flation \cite{nflation}, axion monodromy inflation\cite{monodromy} are the various variant of this natural inflation model where there exist multiple fields with sub-Planckian values of $f$. Multiple axionc fields are aligned in such a way that the effective axion decay constant  could be sub-Planckian and make the model cosmologically viable. Recently proposed Weak-Gravity Conjecture(WGC)\cite{Cheung:2014vva},as well as some string theory construction \cite{banks} put severe constraints on such models of alignment mechanism\cite{wgc1,wgc2}. However, considerations of covariant entropy bound seem to relax the bound\cite{Sloth}. Another way to keep the axion decay constant to have subplanckian value is by introducing a coupling of the Inflaton kinetic term to the Einstein tensor was introduced in \cite{Germani:2010hd,Germani:2011ua}.

In the present paper based on our previous works \cite{debuaxion, modnat}, we point 
out that by introducing a specific form of higher derivative Galileon term for the axion, 
not only we can make inflationary observables compatible with PLANCK, but also we can have sub-Planckian axion decay constant. We will call it as 
G-axion inflation model to make it compatible with the name already exists in the literature.  
However, for sub-Planckian axion decay constant, the price we have to pay is that the axion does not have any oscillatory phase after the inflation. Therefore, the usual reheating mechanism will be inapplicable. Hence we employ instant preheating mechanism proposed by Felder, Kofman and Linde 
\cite{inst} as our rescue. We have also found a bound on $f$ such that our model can be strictly sub-Planckian. For comparison, we will also study models with super-Planckian $f$ \cite{Krippendorf}. We must emphasize at this point that our G-axion inflation scenario with super-planckian axion decay constant turns out be more favorable than the conventional model with regard to the PLANCK observation.

We have organized the paper as follows: following our previous work, we first introduce Galilon type modified axion inflation model. We also discuss in detail their predictions for all the cosmological observables considering some simple choices of kinetic functions. In section-3, we discuss the background dynamics of inflaton during and after the inflation. It can be seen that generically for sub-Planckian axion decay constant inflaton field encounters singularity after some e-folding number. Those models which complete the inflation and become divergent after first zero crossing, we call them non-oscillating model. However for super-Planckian axion decay constant all the models have oscillatory behavior.  
 In the subsequent sections, we will discuss the dynamics after inflation. In section-4, we consider the oscillating model for super-Planckian $f$, and perform the reheating constraint analysis. For concreteness, we also compare our result with the usual axion inflation predictions.  
In section-5, we consider non-oscillating models for sub-Planckian $f$. In order to reheat our universe, we employ the instant preheating mechanism for two different phenomenological coupling of the reheating field with the inflaton and compare the outcomes. In the end, we discuss our results and possible future directions.

\section{Model and its cosmological dynamics}
In the usual inflationary scenario, the action consists of a canonical kinetic term and a 
potential which is sufficiently flat to ensure inflation. However, a non-canonical inflationary model with a higher derivative term in the Lagrangian has special significance from the theoretical point view. One such popular model is Dirac-Born-Infeld (DBI) inflation \cite{dbi1, dbi2, dbi3}, where  non-canonical Kinetic terms also appear. It has been found out that apart from conventional inflationary prediction, these kinds of non-canonical models have other interesting observable predictions such as large non-gaussianity, variable sound speed.
 They are constructed in such a way that it does not lead to any ghosts. A very intriguing such 
class of models exhibiting `Galilean' symmetry ($ \partial_\mu \phi \to \partial_\mu 
\phi + b_\mu $) has been dubbed as Galileon inflation models \cite{galileon1}. 
These Galileon Models has been further extended by Deffayet et. al.\cite{Deffayet}, where the 
Galilean symmetry was broken in order to preserve the second order nature of the equation of motions. A moments pause will ensure us that such a breaking of symmetry is indeed necessary for a viable model of inflation in order to end the inflation and to reheat the universe.
Here we will not be talking about the symmetry breaking mechanism, rather, we will discuss a specific class of inflationary models with a Galileon symmetry broken down to
discrete shift symmetry. 

\subsection{G-axion}
In this section, we will be essentially reviewing the main construction based
on our previous work \cite{debuaxion, modnat}. The Lagrangian 
for G-axion field is  
\bea
S ~=~ \int d^4x \sqrt{-g} \left[\frac {M_p^2}{2}  R - X -  M(\phi) X \Box \phi
- \Lambda^4 \left[1 -\cos \left ( \frac {\phi}{f}\right)\right]\right]
\label{action}
\eea
where $X = \frac 1 2 \pr_{\mu} \phi \pr^{\mu} \phi$ and $\Box = \frac 1 {\sqrt{-g}}\pr_{\mu}(
{\sqrt{-g}}\partial^{\mu})$, and  $\{f,\Lambda\}$ are the axion decay constant and  
axionic shift symmetry breaking scale respectively. $\Lambda$ is also known to be 
associated with the scale of inflation. $M_p = \frac{1}{\sqrt{8 \pi G}}$ is the reduced 
Planck mass. This specific form of the higher derivative term known as
kinetic gravity braiding (KGB)\cite{Deffayet:2010qz}. With the usual FLRW-background ansatz for the spacetime 
\be
ds^2 = -dt^2 + a(t)^2 (dx^2 +dy^2 + dz^2),
\ee
one gets the following Einstein's equations by varying the action with respect to the metric
\bea
3M_p^2 H^2 = - 3H \dot{\phi}^3 M(\phi)-X +  2 X^2 M'(\phi)  +  {\Lambda^4}
 \left[1 -\cos \left ( \frac {\phi}{f}\right)\right]
 \label{E1}
\eea
\bea
M_p^2 \dot{H} = -X\left( 1-3M(\phi) H \dot{\phi} + 
M(\phi)\ddot{\phi}+ M'(\phi)\dot{\phi}^2 \right)
\label{E2}
\eea
and by varying scalar field: 
\bea
\frac 1 {a^3} \frac d {dt} 
\left[a^3\left(1 - {3 H} {M} \dot{\phi} - 2 M' X \right)\dot{\phi}\right]+ \dot{\phi} 
\frac{ d}{ dt}(M'X) + \frac {\Lambda^4}{f} \sin \left ( \frac {\phi}{f}\right) =0.
\label{scalar}
\eea
Where, $H = {\dot{a}}/a$ is the Hubble constant. Following\cite{yokoyama} the slow-roll equation for the axion turns out to be, 
\bea \label{aeq}
3 H \dot{\phi} \left(1 - 3 M(\phi) H \dot{\phi} \right)+ 
\frac {\Lambda^4}{f} \sin \left ( \frac {\phi}{f}\right) =0 .
\label{slowrollscalar}
\eea
From the form of the above equation, we can consider two different ways to inflate our universe. The condition  
$|M(\phi) H \dot{\phi}| \ll 1 $ will give the standard axion inflation scenario, 
while the condition $|M(\phi) H \dot{\phi}| \gg 1 $, provides alternative scenario 
where the higher derivative term comes into play. As emphasized in the introduction, 
the usual canonical axion inflation scenario is tightly constrained from the PLANCK observation. We will see for a wide range of parameter space higher derivative term will play main roll in G-axion-inflation. Solving the slow-roll equation 
of motion for $\dot{\phi}$ we get
\bea
| \dot{\phi} | \simeq M_{p} \left(\frac{V^\prime}{3 M(\phi) V} \right)^{1/2} .
\label{phidot}
\eea  

Using Eq.(\ref{phidot}), the condition for the KGB term to dominate the usual 
slow-roll term can be monitored by defining a parameter
\bea
\tau = M(\phi) V'(\phi) = \frac {M(\phi) \Lambda^4} f \sin \left ( \frac {\phi}{f}\right) \gg 1 .
\label{tau}
\eea 
The slow roll parameters in terms of potential function, denoting $\frac{\phi}{f}$ as 
$\tilde{\phi}$, are
\bea \label{slowroll}
\epsilon &=& \frac{1}{2 \mathcal{A}} \frac{\sin ^{\frac{3}{2}}( \tilde{\phi})}
{ \sqrt{\tilde{M}( \tilde{\phi})} \left(1-\cos ( \tilde{\phi})\right)^2} ~;~
\eta = \frac{1}{2 \mathcal{A}} \frac{\sqrt{\cos ( \tilde{\phi}) \cot ( \tilde{\phi})}}
{ \sqrt{\tilde{M}( \tilde{\phi})} \left(1-\cos ( \tilde{\phi})\right)}  \nno\\
\alpha &=& \frac{2}{ \sqrt{\mathcal{A}}} \frac{\tilde{M}'( \tilde{\phi}) \sqrt{2 \epsilon}}
{ \sqrt[4]{\tilde{M}( \tilde{\phi})^5 \sin ( \tilde{\phi})}} 
~~~;~~~\beta =\frac{1}{36 \mathcal{A}} \frac{\tilde{M}( \tilde{\phi}) 2 \epsilon}
{ \sqrt{\tilde{M}( \tilde{\phi})^3 \sin ( \tilde{\phi})}}  .
\label{slowrollcond}
\eea
We will call $M(\phi)=\frac{1}{s^3}\tilde{M}( \tilde{\phi})$ as the KGB function and $s$ 
is an associated mass scale which will control the strength of the higher derivative term. We define a parameter
$\mathcal{A}=\left( \frac{f^3}{s^3} \frac{\Lambda^4}{M_{p}^4}\right)^{1/2}$ 
which will greatly simplify further calculations. In case of potential driven inflation 
\cite{ohashi}, the KGB function has a significant influence on the inflation 
dynamics. Here our aim is to identify the form of the KGB function based on the principle of constant shift symmetry and try to
satisfy the experimental observation. At this point let us also define an additional 
parameter corresponding to the higher order slow roll parameter which is related 
to another measurable quantity called running of scalar spectral index,
\bea
\xi = M_p^4 \frac{1}{2 M V'} \left(\frac{V''' V'}{V^2}\right) =-\frac{1}{\mathcal{A}^2} 
\frac{\sin (\tilde{\phi })}{\tilde{M}( \tilde{\phi}) \left(1-\cos (\tilde{\phi })\right)^2} .
\eea
For our later convenience, we note the following relation between the KGB function and 
the slow roll parameter,
\bea
\frac{M(\phi) \dot{\phi}^3}{M_{P}^2 H} \simeq -\frac{2}{3} \epsilon .
\label{phidotend}
\eea
Which tells us that, though the higher derivative term dominates the dynamics of slow-roll 
inflation, the standard part of the Lagrangian remains much larger than the Galileon part 
during inflation $(\epsilon \ll 1)$ and they both became comparable only after the end of 
inflation $(\epsilon =1)$.

\subsection{Cosmological quantities: $(n_s, r, dn_s^k)$}
It is worth mentioning that inflation not only solves some of the outstanding problems of 
Big-Bang Cosmology but it is known to be an important mechanism for the generation of seed for the large-scale structure of our universe(see \cite{baumann-tasi} for a review). 
The formation of the this seed of structure formation and their subsequent evolution is known as `The Cosmological Perturbation Theory'(See \cite{cpt1} and \cite{cpt2} for some recent review. The perturbation treatment of Galileon models has been done extensively after the emergence of 
the model \cite{g-cpt, Gao:gcpt}. In this section, we will note down some of the important 
cosmological quantities which are being measured in cosmological experiments. Following 
\cite{yokoyama}, the amplitude of the power spectrum of the curvature perturbation is written as
\bea
P_{\cal R} =  \frac {3 \sqrt{6}}{64 \pi^2}\frac {H^2}{M_p^2 \epsilon} .
\label{PR}
\eea
The spectral tilt and its running can be easily found out using the relation, 
$\frac{d}{d \hspace{0.1cm} lnk} = \frac{\dot{\phi}}{H}\frac{d}{d\phi}$
\footnote{In case of KGB inflation,the slow-roll scalar field equation can be used to 
find $\frac{\dot{\phi}}{H} = -\frac{M_p^2}{V(\phi)}\left(\frac{V'(\phi)}{M(\phi)}\right)^{1/2}$ }
\bea
n_s - 1\equiv \frac{d\hspace{0.1cm} ln P_{\cal R}}{d \hspace{0.1cm} lnk}=- 6 \epsilon + 
3 \eta  - \alpha , \\ \nno \frac{dn_s}{d\hspace{0.1cm} lnk} \equiv dn_s^k= -3 
\xi +24\epsilon \eta -24 \epsilon^2 -3 \alpha^2 - 8 \alpha \epsilon + 4 \alpha \eta +
3 \eta^2 + 18 \beta .
\label{ns}
\eea
The power spectrum and the spectral index of the primordial gravitational wave are:
\bea
P_{\cal T} = \frac{8}{M_p^2}\left(\frac{H}{2\pi}\right)^2~~;~~n_T &=& - 2 \epsilon .
\label{PT}
\eea
Another important quantity of cosmological importance is the tensor to scalar ratio
\bea
r =  - \frac {32 \sqrt{6}}{9} n_T
\label{r}
\eea
The idea of inflation was introduced in the first place to solve the horizon problem 
and the flatness problem. It has been commonly said that we need sufficient amount of 
inflation to solve the aforementioned two problems. This idea of `sufficient' inflation is quantified 
by what is known as the `e-folding' number which will be described next.


\subsection{Number of e-folds and modified Lyth bound} 
The `e-folding' number is given by by the following expression,
\bea
{\cal N}= \int_{t_1}^{t_2} H dt  =\int_{t_1}^{t_2} \frac {H}{\dot{\phi}} d\phi = \frac 1 {M_p} 
\int_{\phi_{end}}^{\phi_{in}} \frac {\tau^{\frac 1 4}} {\sqrt{2 \epsilon}} d\phi.
\label{efold}
\eea
From the cosmological observation, the e-folding number is found to be ${\cal N} \geq 50 $.
It is apparent that the amount of inflation must be proportional to the amount of field 
excursion during the slow roll inflation. This quantity is known as Lyth bound 
\cite{lyth}(also see \cite{easther,baumann}). Lyth bound is assumed 
to play an important role in constraining the parameter space of a model under consideration
from the low energy effective field theory point of view. In the slow roll approximation, one
can compute the amount of field excursion by series expansion in terms of e-folding number 
calculated from the onset of inflation as follows
\bea
\phi(\delta{ \cal N}) = \phi_{in} + \frac{\partial \phi}{\partial {\cal N}} \delta{\cal N} + 
\frac 1 2 \frac{\partial^2 \phi}{\partial {\cal N}^2} \delta{\cal N}^2 + \dots .
\eea
Therefore, up to second order in slow roll, one can write down the general expression
for the amount of field excursion $\Delta \phi$ for e-folding number $\delta {\cal N}$ as
\bea
\Delta \phi = |\phi(\delta{\cal N}) - \phi_{in}| = \delta {\cal N}
 \left(\frac{\partial \phi}{\partial {\cal N}}\right) \left[1 + \frac{\delta {\cal N}}{2} 
\left(2 \epsilon - \eta + \frac{\alpha}{2}\right)\right] .
\eea
Therefore, by using eq.(\ref{efold}), we get 
\bea
\Delta \phi = \delta {\cal N}
\left|\frac {\sqrt{2 \epsilon}}  {\tau(\phi)^{\frac 1 4}}\right| \left[1 + \delta {\cal N} 
\left(2 \epsilon - \eta + \frac{\alpha}{2}\right)\right] .
\eea

From the above expression for the e-folding number 
assuming, that the slow-roll parameters and $\tau$ behave monotonously, we can write
\bea
{\cal N} \lesssim \frac {\Delta \phi} {M_p} 
\left|\frac {\tau(\phi)^{\frac 1 4}} {\sqrt{2 \epsilon}}\right|_{max} = \frac {\Delta \phi} {M_p} 
\left|\frac {\tau_{max}^{\frac 1 4}} {\sqrt{2 \epsilon_{min}}}\right| .
\eea
In deriving the above expression, we have kept only the leading order term in slow roll. However, above expression can get significant correction near the end of inflation where slow roll parameters are not small. We will do the detailed study on this issue and its consequence on the model for our future publication. 
Now, Eqs(\ref{PR}-\ref{r}) can be used to relate the field excursion during inflation 
with the scalar-to-tensor ratio as,
\bea
{\Delta \phi} \gtrsim ~({\cal N} {M_p}) \left|\frac {\sqrt{2 \epsilon_{min}}}  {\tau_{max}^{\frac 1 4}}\right| = 
\frac f {\sqrt{{\cal A}}} \frac {{\cal N}}{{\cal T}_{max}} \sqrt{\frac { 9 r} { 36 \sqrt{6}}},
\label{deltaphi}
\eea
where, 
\bea
{\cal T}_{max}= (s^3 M(\tilde{\phi}_{in}) \sin \tilde{\phi}_{in})^{\frac 1 4}. \nno
\eea
As we can see there is an important difference in the above expression for Lyth Bound with 
usual slow-roll inflation case via the presence of the term $\tau(\phi)$ which contains the 
KGB term $M(\phi)$. This feature will help us to construct a model 
with detectable tensor-to-scalar ratio with sub-planckian axion decay constant. Therefore,
we are in a position to compare the Lyth bound between G-axion inflation and the usual canonical 
slow-roll inflation, which can be written in a model independent way as 
$\Delta\phi_{slow-roll} \sim {\cal N} \sqrt{\frac{r}{8}}$. For, $r=0.08$ and ${\cal N}=60$, $\Delta \phi_{slow-roll} \gtrsim 6$. 
Now since we have freedom in choosing $f$, the modified Lyth bound can be made very small. 
As an example, for $\mathcal{A} = 94$, if we choose $f = 0.5 M_p$, one gets
$\Delta \phi_{KGB} \gtrsim 0.59$. 
In the subsequent section, we will consider some simple phenomenological form of the KGB functions keeping the constant shift symmetry of the Lagrangian intact, and construct both  super and sub-Planckain inflation models in compatible with PLANCK.

\subsection{Simple choices of $M(\phi)$ and determination of cosmological parameters}
\begin{figure}[t]
\centering
\includegraphics[width=\textwidth]{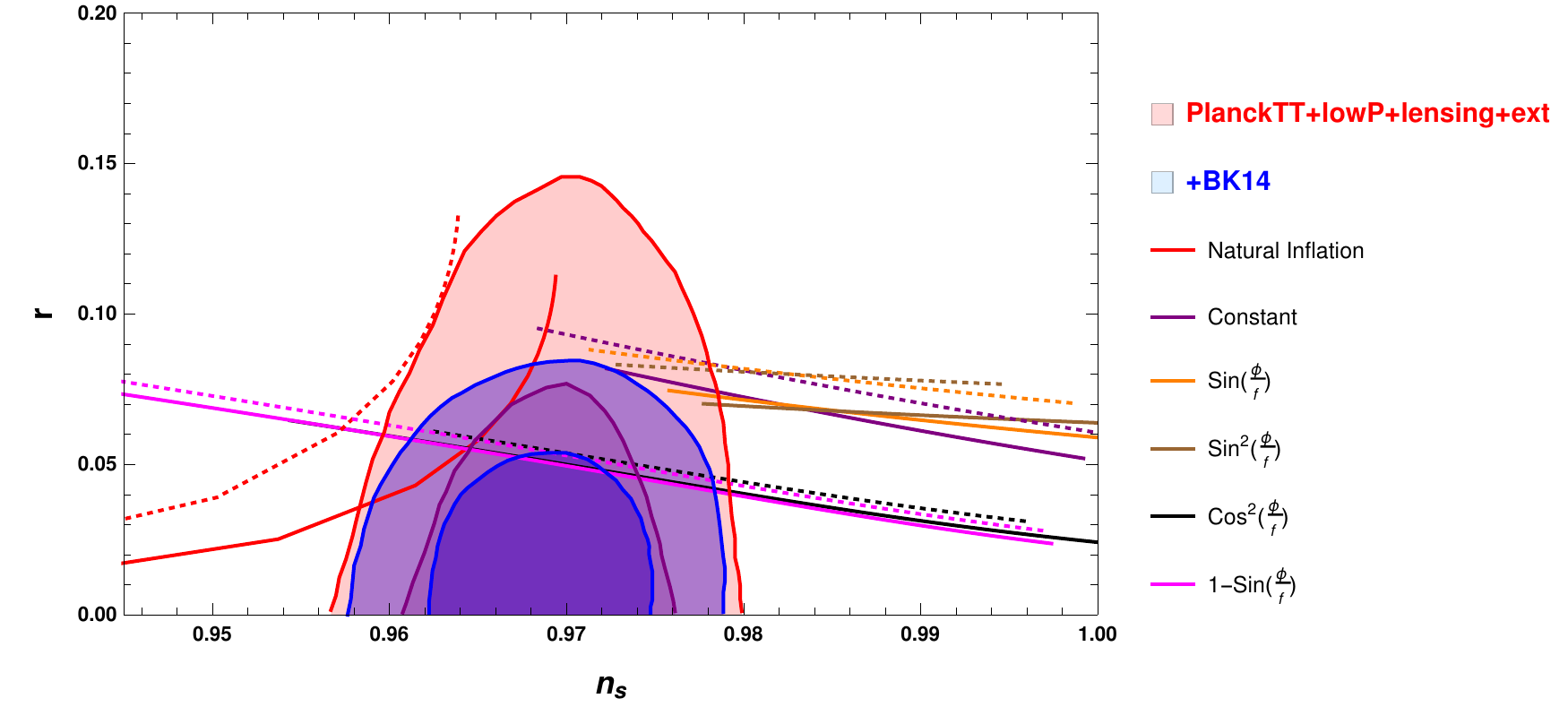}
\caption{ \scriptsize 
	$n_s$ vs $r$ plot for five different forms of $ M(\phi) $ for ${\cal N}= 55$ 
(dotted lines), and ${\cal N} = 65$ (solid lines). For comparison, red curves are plotted for usual axion inflation.}
\label{planckplot}
\end{figure}

\begin{figure}[t!]
\centering
\includegraphics[scale=0.5]{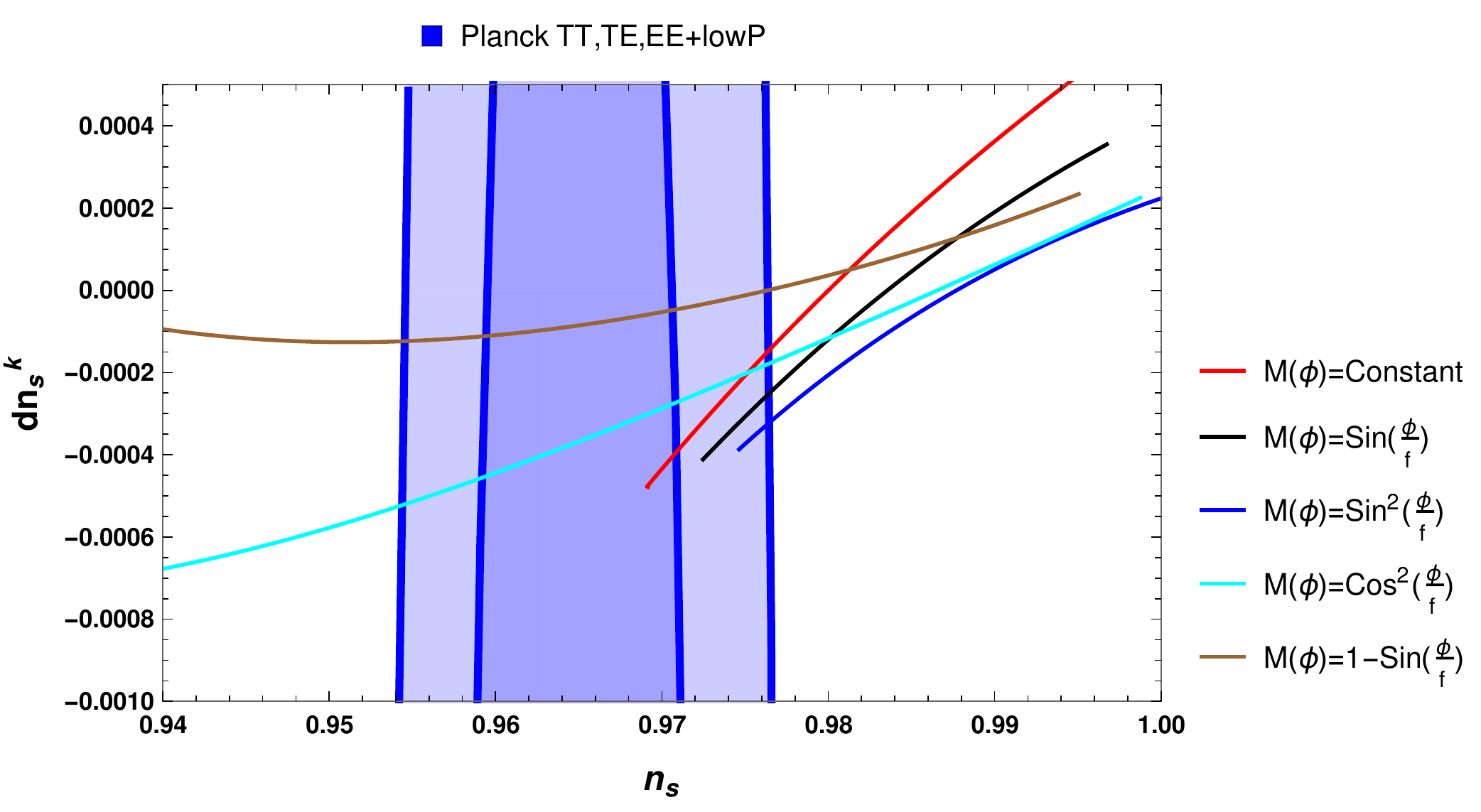}
\caption{\scriptsize $ dn_s^k$ vs $n_s$ plotted for five different forms of $ M(\phi) $ for ${\cal N}= 60$ 
e-folding on the background of Planck TT,TE,EE+low P data $(dn_s^k=- 0.003 \pm 0.007)$}
\label{running}
\end{figure}

As we have emphasized in the beginning, we will be considering some simple
functional form of $M(\phi)$ exhibiting shift symmetry of the 
axion field. As we will see from our analysis, the CMB observables are not sufficient to constrain the value of the axion decay constant for the model under consideration. Therefore, the dynamics after the inflation will be important and we will indeed see the existence of a critical value of axion decay constant $f_c$ which separates super and sub-Planckian scenarios in terms of oscillation dynamics.  
However, in this section, we will mainly compute the 
inflationary observables and compare our result with the current observational bounds coming from
PLANCK for some simple choices of $M(\phi)$. We consider the following forms of $M(\phi)$:
\begin{enumerate}
\item[I]: $M(\phi) = \text{Constant}$,
\item[II]: $M(\phi) = \sin(\phi/f)$, 
\item[III]: $M(\phi) = \sin^2(\phi/f)$,
\item[IV]: $M(\phi) = \cos^2(\phi/f)$,
\item[V]: $M(\phi) = 1-\sin(\phi/f)$,
\label{mset}
\end{enumerate}
The CMB observables with the above choices of function have been presented in figs.(\ref{planckplot}) and (\ref{running}). For comparison, we also plotted the standard axion inflation model. From our analysis and also clear from the $(n_s,r)$ plot,
we have two different categories of model based on their predictions. The best fit models are of type $(IV,V)$. One can clearly see that within $1\sigma$ region of $n_s$, natural inflation model is almost ruled out. On the other hand the our aforementioned best fit model predicts as low as $r \simeq 0.04$ within the same $n_s$ region. 
 Other three types $(I,II,III)$ are marginally fitting with the data.
All these models predict very high value of tensor to scalar ratio $r$ which can be measured in the near future CMB experiments. Particularly, for $M(\phi) = constant$, which is the best out of this group predicts, $r \simeq 0.07, 
n_s = 0.973$ for $N=65$ which is within $2 \sigma$ region 
of CMB data.
Interestingly, as we will discuss in detail, those two types of models show distinct behavior after the inflation depending on the value of axion decay constant.     

All the inflationary prediction will only depend on the emergent 
parameter ${\cal A}$. Therefore, we need further condition to constrain 
the value of $(f,s)$. We determine the height of the axion potential 
from the WMAP normalization for the scalar power spectrum \cite{komatsu}
\bea \label{pr}
P_{\cal R} =  \frac {{\cal A}\sqrt{6 }}{32 \pi^2} 
\left( \frac {\Lambda}{M_p}\right)^4 
\frac{(1-\cos \tilde{\phi}_1)^3 \sqrt{ s^3 M(\tilde{\phi}_1)}}{\sin ^{\frac{3}{2}} \tilde{\phi}_1} \simeq 2.4\times 10^{-9},\nno.
\eea
The value of $\Lambda$ turned out to be $10^{-2} M_p$ for all the above models under consideration.
Therefore, by using the values of $({\cal A}, \Lambda)$, the following expression of $\mathcal{A}$,
\bea \label{ratio}
\frac {s^3}{f^3} = \frac 1 {{\cal A}^2} \frac {\Lambda^4}{M_p^4} ,
\eea
will provide us all possible values from super-Planckian to sub-Planckian for $(f,s)$. From particle physics, we can set the lower bound on $f$ as a $U(1)$ symmetry breaking scale which should be higher than the inflationary scale implying $f > \Lambda$. However, the bound on the values of $s$ from the any fundamental theory is not evident to us and obviously needs further study. Hence, we chose $s$ as a free parameter which measures the strength of the higher derivative operator. Up to this point, we have all the necessary ingredients for a successful inflation producing the correct values of the scalar spectral index within the observational limit. Our next task would be to go beyond and study the axion dynamics after inflation when the particle production from inflaton field will be initiated. At this point let us re-emphasize the fact that during the inflation the change in axion field turned out to be 
\bea
\Delta \phi  = (\tilde{\phi}_1 - \tilde{\phi}_2) \times f  <  f .
\eea
Therefore, for a wide range of axion decay constant $f$, we will have  sub-planckian field excursion with the required value of e-folding number. 
Consequently our model will not have any serious trans-planckian problem. 

\section{Evolution of the scalar field: sub-Planckian $(f,s)$}

Before we go for cosmological dynamics, let us first try to understand the region of stability of our G-axion system. Even though we do not have more than two time derivative in the equation of motion, higher derivative term in the Lagrangian generically gives rise to various pathological behavior for the fluctuation dynamics in a non-trivial background. Due to those behaviors, such as propagating ghost, gradient instability, superluminal speed of propagation, the effective field theory under consideration may not have conventional UV complete description. For completeness and also in order to have qualitative understanding of the aforementioned pathologies we will apply the well known method of characteristics \cite{Deffayet, Easson:2011zy,Easson:2013bda}. In this method the procedure is the following. From the action given in
\ref{action}, the axion field equation can be written as
\bea
P^{\mu\nu}\nabla_{\mu}\nabla_{\nu} \phi +Q^{\mu\nu\alpha\beta} (\nabla_{\alpha}\nabla_{\beta} \phi)(  \nabla_{\mu}\nabla_{\nu} \phi )- 2 X M''(\phi) -V'(\phi) - 2 X M(\phi) (2 X-4 X^2 M(\phi))  =0
\eea
where, we have 
\bea
&&P_{\mu\nu} = g_{\mu\nu} - 2 M'(\phi)\nabla_{\mu}\phi \nabla_{\nu} \phi - 2 X M(\phi)^2( X g_{\mu\nu} - 2 \nabla_{\mu}\phi \nabla_{\nu} \phi) \\ &&Q^{\mu\nu\alpha\beta} =  {M(\phi)} g^{\mu\nu} g^{\alpha\beta} -\frac {M(\phi)}{2}\left(g^{\alpha\mu}g^{\beta\nu} + g^{\alpha\nu}g^{\beta\mu}\right)
\eea 
Now information regarding the stability and the other aforementioned pathological behaviors of the background solutions are generically encoded in an effective metric on which the fluctuation propagates. Considering the liner order fluctuation of the above equation for the axion, one finds
\bea
(P^{\mu\nu} + 2 Q^{\mu\nu\alpha\beta} \nabla_{\alpha}\nabla_{\beta} \phi)  \nabla_{\mu}\nabla_{\nu} \delta\phi + \cdots  =0 .
\eea
Where ``$\cdots$" corresponds to all the terms which does not contain second derivative on $\delta \phi$. From the above equation for the fluctuation the aforementioned effective metric can be identified as the coefficient of $\nabla_{\mu}\nabla_{\nu} \delta\phi$, which is
\bea
{\cal G}^{\mu\nu} = (P^{\mu\nu} + 2 Q^{\mu\nu\alpha\beta} \nabla_{\alpha}\nabla_{\beta} \phi). 
\eea
\begin{figure}[t!]
	\centering
	\includegraphics[scale=0.5]{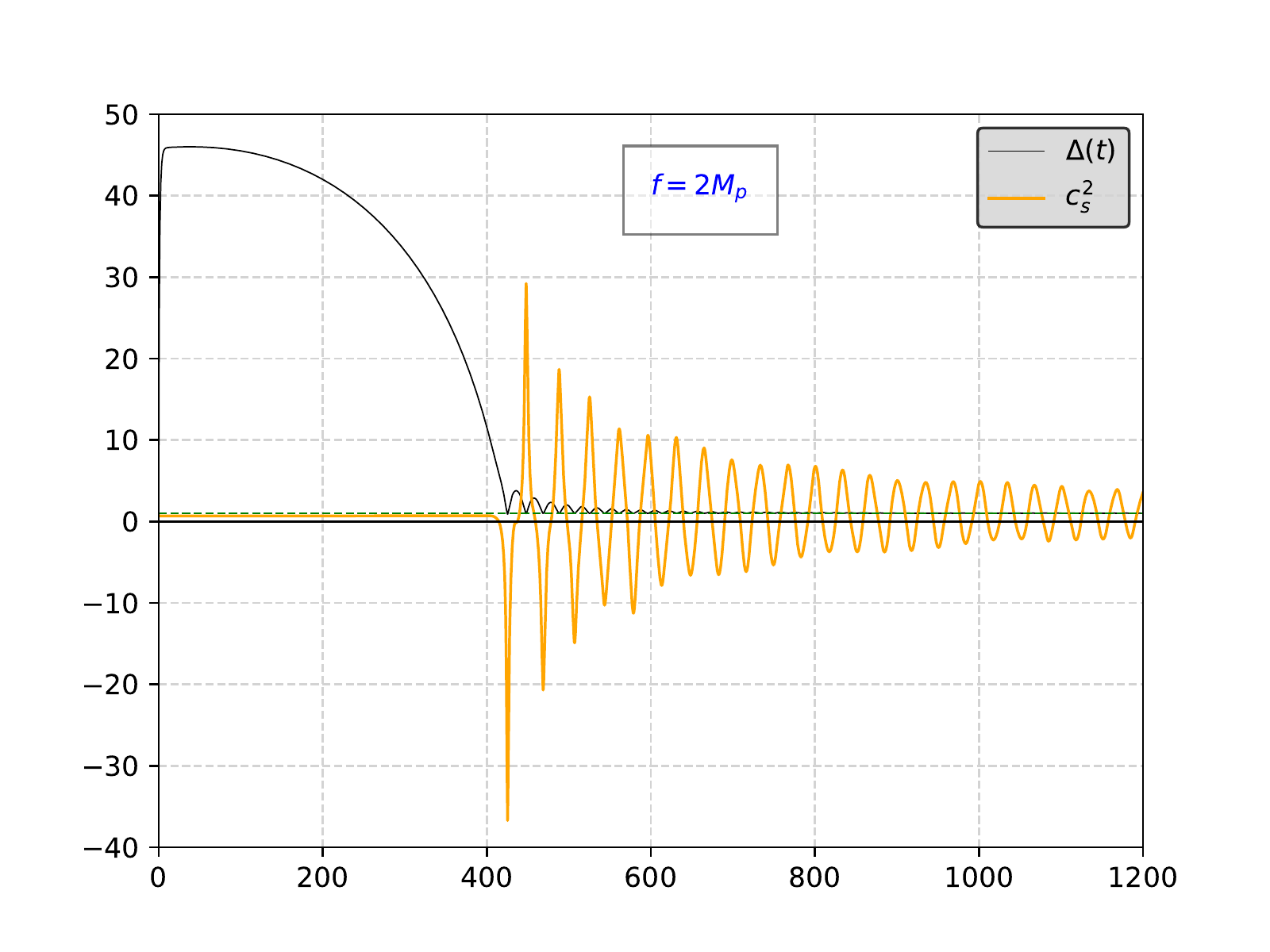}\hspace*{1cm}
	\includegraphics[scale=0.5]{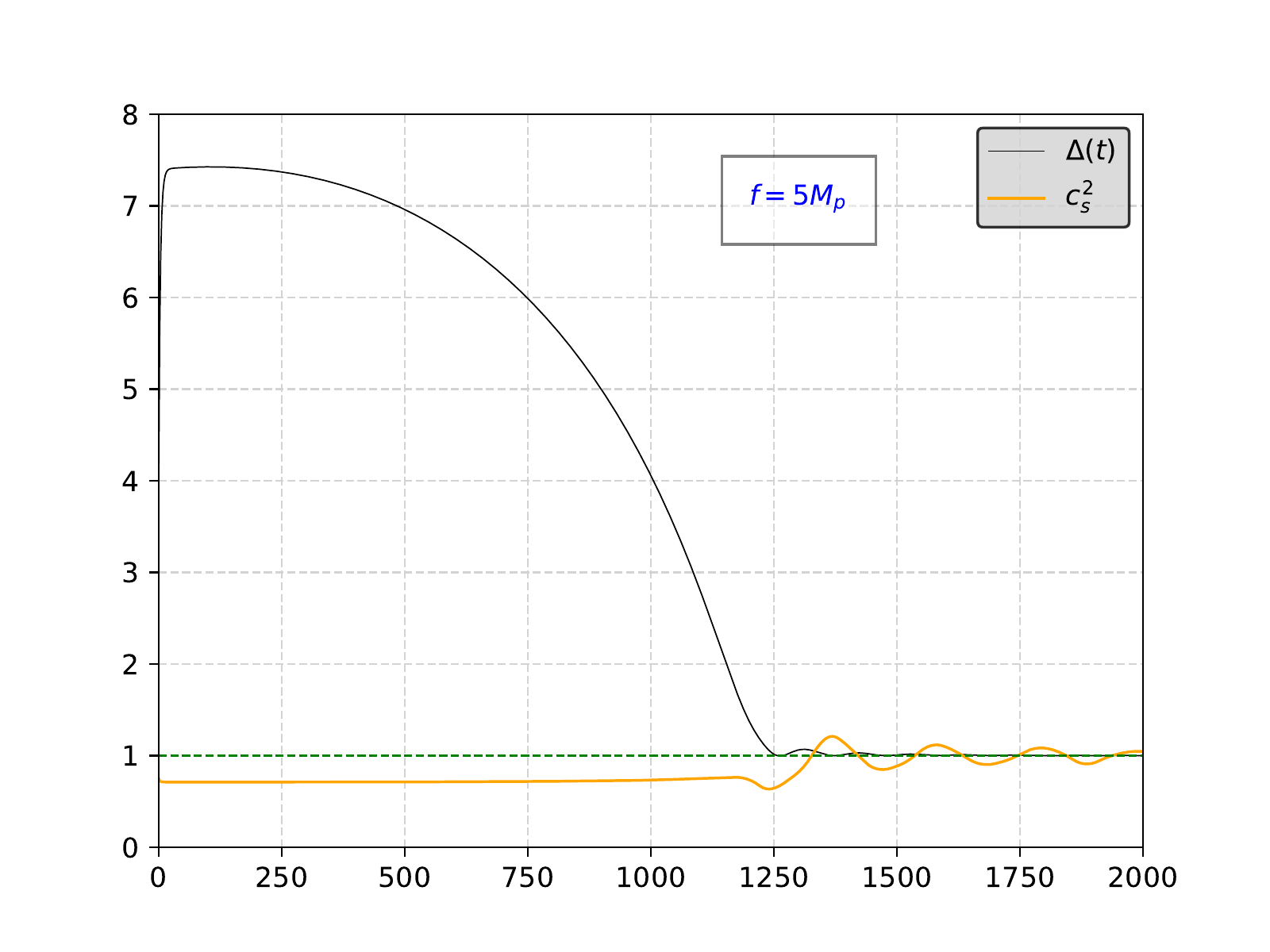}\\
	\includegraphics[scale=0.5]{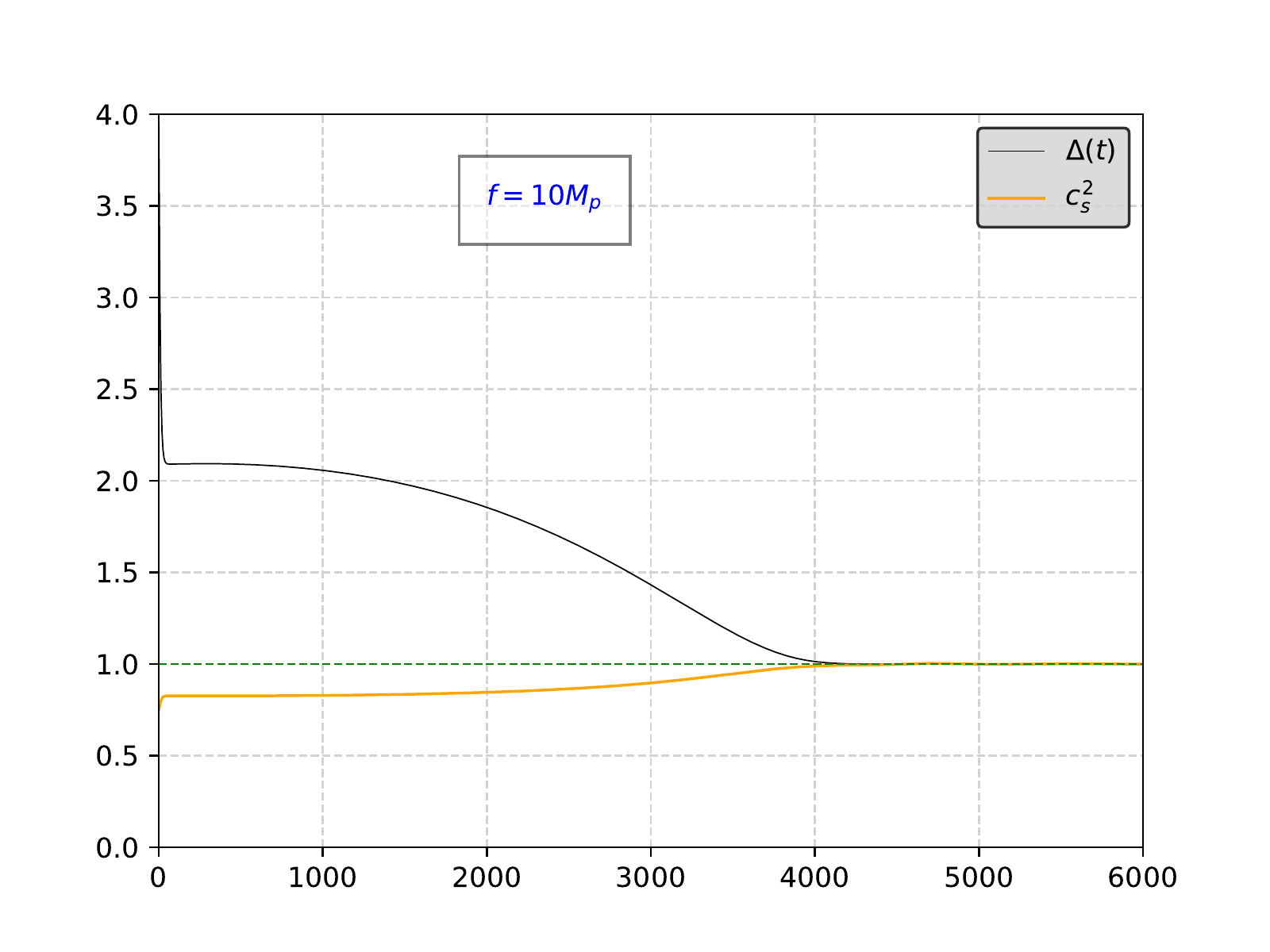}\hspace*{1cm}
	\caption{\scriptsize The evolution of $(\Delta(t), c_s^2)$ with time for three different values of the axion decay 
		constant for $M(\phi)=Sin(\phi/f)$ when $\mathcal{A}=92$. The value of $c_s^2$ is always smaller than unity during inflation. While the parameter $\Delta(t)$ is positive for all $f < f_c$ ensuring stability.} 
	\label{denominator}
\end{figure}
This metric is calculated in a given background solution for the axion field such as inflationary background for the present case. Therefore, for any generic time dependent  background, the effective metric components take the following form,
\bea
{\cal G}^{00} &=&-\left(1 - 6 H M(\phi) \phi ' +2 \phi '^2 M'(\phi )+\frac{3}{2} M(\phi)^2 \phi '^4 \right) \\
{\cal G}^{ij} &=& \frac {\delta^{ij}}{a^2} \left(1 - 4 H M(\phi) \phi ' - 2 M(\phi) \phi '' -  \frac 1 2 M(\phi)^2 \phi '^4 \right) .  
\eea

For stability of the system, following conditions has to be satisfied, $-{\cal G}^{00} = \Delta(t) > 0$ and the associated effective sound speed,
\bea
 c_s^2 = \frac {\left(1 - 4 H M(\phi) \phi ' - 2 M(\phi) \phi '' -  \frac 1 2 M(\phi)^2 \phi '^4 \right)}{\left(1 - 6 H M(\phi) \phi ' +2 \phi '^2 M'(\phi )+\frac{3}{2} M(\phi)^2 \phi '^4 \right)} \leq 1
\eea
For illustration we have plotted those parameters $(\Delta(t), c_s^2)$ in fig.\ref{denominator} for a particular cosmological model with $M(\phi) = \sin(\phi/f)$. We can clearly see that depending upon the axion decay constant, before the $\Delta(t)$ parameter could cross zero, the superluminal velocity arises. If we further decrease the value of $f$, the sound speed becomes imaginary. Although for larger values of the axion decay constant these instabilities could be avoided, we are mainly interested in subplanckian values of $f$. Hence, a further modifications to our model are necessary. Interestingly a possible resolution to these instabilities due to sound speed has been proposed recently\cite{Easson:2018qgr} in the context of effective field theory of cosmological perturbations. Implication of these ideas could be interesting to explore further. This discussion provides us a hint for the existence of a critical value of axion decay constant below which standard evolution will not be possible after the end of inflation. In our subsequent discussion we see how the $\Delta(t)$ parameter will effect the evolution of background dynamics of $\phi$. The detail fluctuation analysis we left for our future studies. However, it is important to mention that all our models under consideration, the fluctuation are well behaved during inflation. Hence, inflationary observables will not be effected by those instability even if the value of $f$ is sub-Planckian.      

It must be clear from previous discussions that the value of $(f,s)$  will be constrained from appropriate slow roll condition. For our G-axion or more generally Galileon inflation models, the nature of initial slow roll condition is dependent upon the choice of $f$ as it has to satisfy a non-trivial relation (\ref{tau}). In this section we will try to understand, qualitatively, the evolution of the scalar field depending on the initial conditions as well as our model parameters.   
The equation of motion for the scalar field (\ref{scalar}) combined with eq.(\ref{E2}, in unit of $M_p=1$) can be written as
\begin{align}
\Delta(t)\ddot{\phi}(t) + 
\left[3H -9M(\phi)H^2 \dot{\phi} +\frac{1}{2}M''(\phi) \dot{\phi}^3 - \frac{9}{2}M^{2}(\phi)H^2 \dot{\phi}^4 + \frac{3}{2} M(\phi) M'(\phi) \dot{\phi}^5 \right] \dot{\phi}(t) 
 + V'(\phi)=0
\label{singular}
\end{align}
From the above equation, it is evident that the solution may encounter
singular behavior depending upon the initial condition when the coefficient $\Delta(t)$ becomes zero. The initial conditions in turn depend upon the parameters $(f,\Lambda,s)$. In our numerical calculation, it has been found that the scalar field does show singularity or oscillatory 
behavior depending on the values of the above mentioned parameters. The variation of $\Delta(t)$ and $c_s^2$ with time for three different
values of $f$ has been shown in the fig.(\ref{denominator}).

\begin{figure}[t!]
\centering
\includegraphics[scale=0.5]{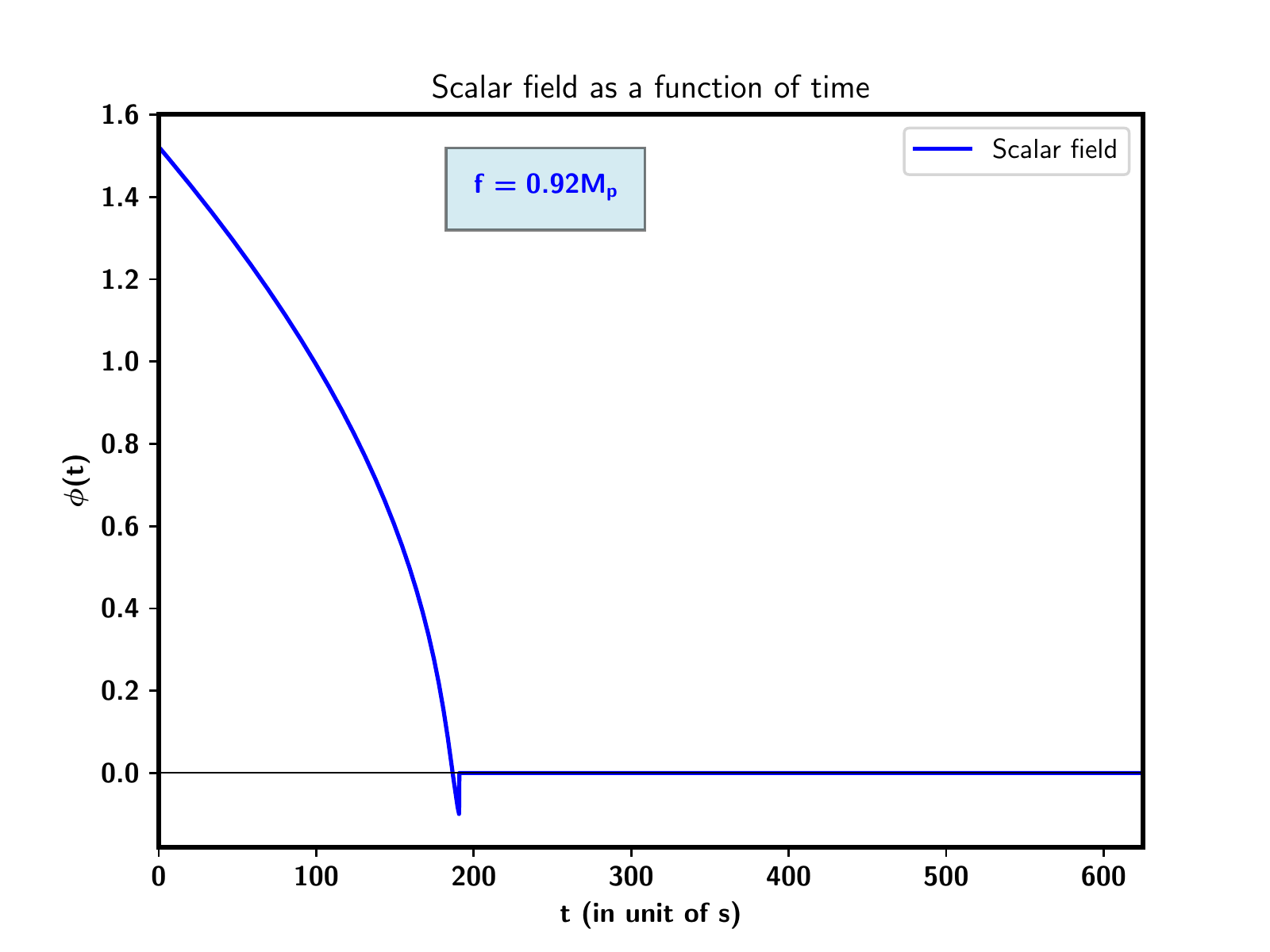}\hspace*{1cm}
 \includegraphics[scale=0.5]{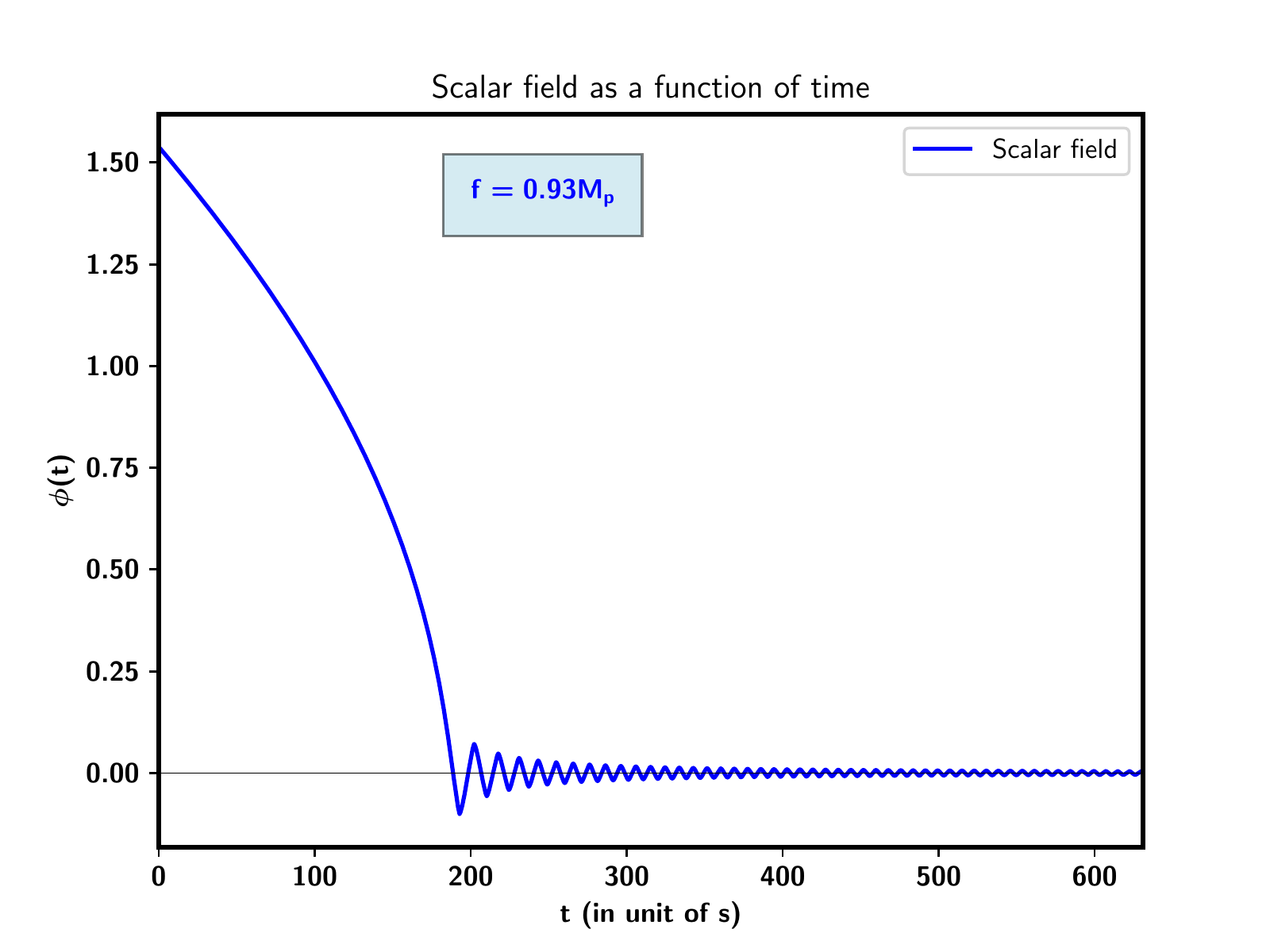}
\caption{\scriptsize The evolution of the scalar field for two values of f when the solution transits from oscillatory to singular}
\label{inflatonevlb}
\end{figure}

It has been found that the minimum value of the coefficient tends to zero with decreasing $f$. The smallest value of axion decay constant which yields an oscillatory solution is when the coefficient is about to touch the zero axis. These numerical results have been confirmed in four different numerical environments. Fig.(\ref{inflatonevlb}) shows an illustration of the scalar field
 behavior for $M(\phi) = Sin(\phi/f)$. We have shown the behavior of the scalar field solution for two adjacent values of $f$ in unit of Planck when the scalar field solution makes a transition from being singular to oscillatory. For all different functional form of $M(\phi)$, we found similar behavior for different critical values of $f$. However, it is interesting to note that models of type $(I,II,III)$, which are marginally fitting with the PLANCK observation, always complete the inflation before the field dynamics becomes singular for sub-Planckian decay constant. On the other hand best fit models of type $(IV,V)$ do not show such behavior as singularity appears well before the completion of inflation for sub-Planckian $f$. 
This behavior can be inferred also from the fact that for aforementioned models the slow roll condition is violated well before the inflation ends for sub-Planckian $f$. 
To this end we emphasize the following observation: all the models under consideration will have coherent oscillation for super-Planckian axion decay constant.
Two of them can explain CMB observation for sub-Planckian $f$.
From our numerical analysis, we found $f$ can even take super sub-Planckian value $\ll M_p$ without changing the properties of the solution. However, from the effective field theory point view,
the value of axion decay constant should be limited to ${\cal O}(1) > f > \Lambda \simeq 10^{-2}$. None the less, the price we pay for those sub-Planckian model is that after inflation the inflaton field shows singular behavior, we call them non-oscillating axion models. Hence, as will be subsequently discussed, for those models we will employ the instant preheating mechanism to reheat the universe. The appreance of this type of singularity is related to the pressure singularity in KGB models. It has been shown in \cite{Deffayet:2010qz} that this pressure singularity implies infinite curvature scalar by a finite value of Hubble parameter and a divergent sound speed.
 
An interesting point about the solution is that there exists a minimum value of the field excursion, $\Delta\phi_c$ above which we have the oscillatory solution. This value is independent of the parameters $(f,~ \Lambda~ \text{and}, s)$ for a particular form of $M(\phi)$. In the table \ref{tab_oss}, we have shown the critical value of $f$ and the minimum field excursion when the oscillatory behavior sets in.


\begin{figure}[t!]
\label{avsdensityplot}
	\centering
	\includegraphics[width=6.00cm,height=4.00cm]{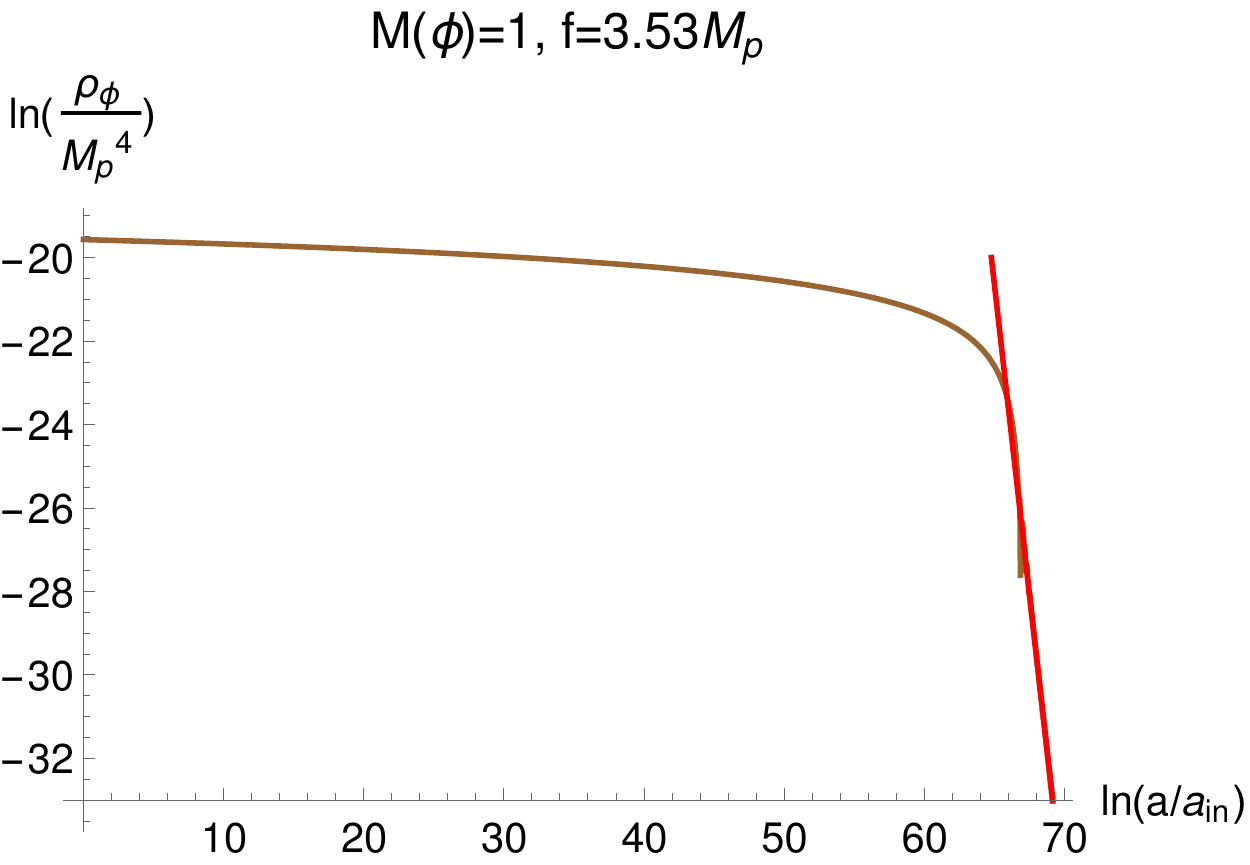}\hspace*{1cm}
	\includegraphics[width=6.00cm,height=4.00cm]{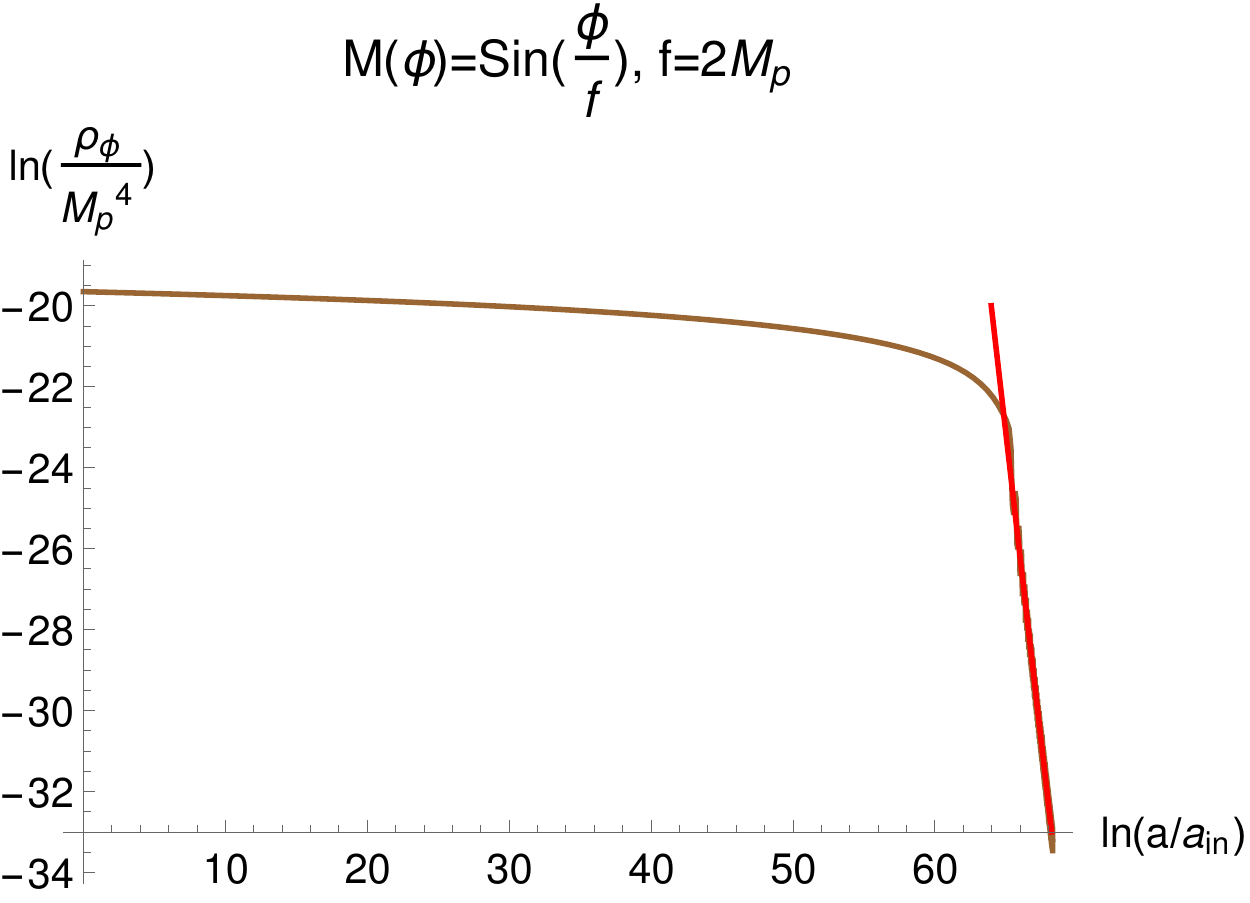}
	\caption{\scriptsize Evolution of energy density of the inflaton field for two different oscillating G-axion models. The fitting red line shows the behavior of $\rho_{\phi} \sim a^{-3}$.}
\end{figure}
\begin{table}
	\begin{center}
		\begin{tabular}{ |p{2.0cm}||p{0.5cm}|p{2.5cm}|p{1.1cm}|  }
			\hline
			\multicolumn{4}{|c|}{Summary of the scalar-field dynamics} \\
			\hline
			$M(\phi)$ &$\mathcal{A}$ &$\Delta \phi_c (=\phi_1-\phi_2)$  &$f_c/M_p$  \\
			\hline
			\hline
			Constant & 76   &  4.85158 & 3.24 \\
					
			\cline{2-4}
			& 100   & 4.86432 & 3.62 \\
			
			\cline{2-4}
			\hline 
			$Sin(\phi/f)$	& 92  & 1.27722 & 0.93 \\

			\cline{2-4}
			& 150   &  1.26267 & 1.1 \\
			
			\cline{2-4}
			\hline
		$Sin^2(\phi/f)$	& 88  & 3.30 & 2.5 \\
			
						\cline{2-4}
			& 95   &  3.28 & 2.55 \\
			\cline{2-4}
			\hline
		\end{tabular}
		\caption {\scriptsize $f_c$ is the minimum value of $f$ for which the oscillatory behavior set off. The values of ${\cal A}$ are so chosen that the predicted value of  $n_s$ is the central value of PLANCK.}
		\label{tab_oss}
	\end{center}
\end{table}
The PLANCK-2015 observation suggests the value of $\mathcal{A}$ should be  within $40 \sim 360$, considering different models under consideration. Using the condition for dominating higher derivative term during inflation eq.(\ref{tau}), aforementioned range of ${\cal A}$ constrains the value of axion decay constant to be
$f \ll 6M_p \sim 18 M_p $.

It is now clear that for a wide range of axion decay constant our G-axion model fits extremely well with the CMB observation. For further understanding, it is of prime importance to go beyond the inflationary dynamics. However, for the singular behavior of the inflaton field which we have been discussing in the previous sections, we need a detailed analysis considering the additional reheating field coupled with the inflaton. Thanks to the instant preheating mechanism \cite{inst} which has been proposed based on the mechanism of parametric resonance\cite{kofman}.
Instead of considering the full dynamics, for non-oscillatory model we employ the aforementioned instant preheating mechanism which acts at the instant of the first zero crossing of the inflaton field \cite{campos,panda_sami,gr-qc-0307068,sami_sahni}.  
 However, for the oscillatory solution, the usual treatment of parametric resonance as well as perturbative reheating by solving the appropriate Boltzmann equation\cite{ckr1, gkr} must be applicable. For those oscillatory models, we compute the behavior of density and pressure of the inflaton field  in terms of background expansion, 
\bea
\rho_{\phi} = \frac{1}{2}\dot{\phi}^2 -3 M(\phi) H \dot{\phi}^3 + \frac{1}{2}M'(\phi)^4 +
 \Lambda^4 \left[ 1- \cos\left(\frac{\phi}{f}\right)\right] \\
P_{\phi} = \frac{1}{2}\dot{\phi}^2 + \frac{1}{2}M'(\phi)^4 + M(\phi)\dot{\phi}^2 \ddot{\phi} - 
\Lambda^4 \left[ 1- \cos\left(\frac{\phi}{f}\right)\right].
\eea
As expected, by using the numerical fitting shown in figure(\ref{avsdensityplot}), we found $\rho_{\phi}\propto 1/a^{3}$, with the equation state $\omega =0$.  Thus inflaton behaves 
as a matter(pressure-less dust) field during the coherent oscillating phase.

As emphasized earlier, in the subsequent sections our main goal is to study the dynamics of G-axion model during reheating phase for both oscillatory and non-oscillatory cases. For the oscillating axion model, we do the general reheating constraint analysis and analyze the reheating temperature and duration of perturbative reheating considering the constraints from CMB. Finally for the non-oscillatory model we employ instant reheating scenario to reheat the universe.

 \section{Oscillating axion: Constraints from rehating predictions}
 As we have seen, depending upon the value of axion decay constant we have two different possibilities for the axion field dynamics after the inflation. In this section, we consider studying the reheating constraints for the oscillating models based on the analysis of \cite{kamionkowski}. The evolution of cosmological scales throughout the evolution history of our universe, and the entropy conservation \cite{liddle,kamionkowski,cook} have been proved to be an important way to constraining the inflationary model. One of the important steps towards this is to parametrize the reheating phase by its characteristic reheating temperature $(T_{re})$, the equation of state parameter $(\omega_{re})$, and reheating e-folding number $(N_{re})$. Following our  previous work \cite{debuGB}, aforementioned parameters are inter-related through the following equations, 
 \begin{eqnarray} \label{nretre}
 && N_{re} = \frac{4(1+\gamma)}{(1-3\omega_{re1})+\gamma(1-3 \omega_{re2})}\left[61.6 - \ln\left(\frac{V_{end}^\frac{1}{4}}{H_k}\right) -N_k\right]\\
 && T_{re} = \left[\left(\frac{43}{11g_{re}}\right)^\frac{1}{3} \frac{a_0 T_0}{k} H_ke^{-N_k}\right]^{\frac{3[(1+\omega_{re1})+\gamma(1+\omega_{re2})]}
 	{(3\omega_{re1}-1)+\gamma(3 \omega_{re2}-1)}}
 \left[\frac{3^2.5V_{end}}{\pi^2g_{re}}\right]^{\frac{1+\gamma}{(1-3\omega_{re1})+\gamma(1-3 \omega_{re2})}} .
 \end{eqnarray}
 In the above expressions we have considered two stage reheating process as will be subsequently explained. In this two stage reheating process, we parametrize the reheating parameters as $(N_{re}=N^1_{re}+N^2_{re},T_{re},\omega^1_{re},\omega^2_{re})$. 
 Where, $N^1_{re}, N^2_{re}$ are efolding number during the first and second stage of the 
 reheating phases with the equation of state parameters $\omega^1_{re}, \omega^2_{re}$ 
 respectively. In the derivation of above two formulas, we also assume the change of reheating stage from the first to the second one to be instantaneous.
 
 With all the aforementioned ingredients we study the possible constraints on our G-axion inflation model. As we have seen from the previous analysis, all the cosmological quantities during inflation can be expressed in terms of the emergent 
 parameter ${\cal A}$. Therefore, we will have a range of value of ${\cal A}$ for which cosmological predictions will be within the range of observed values of $(n_s, r)$. 
 \begin{figure}
 	\begin{center}
 		\includegraphics[width=005.0cm,height=04.0cm]{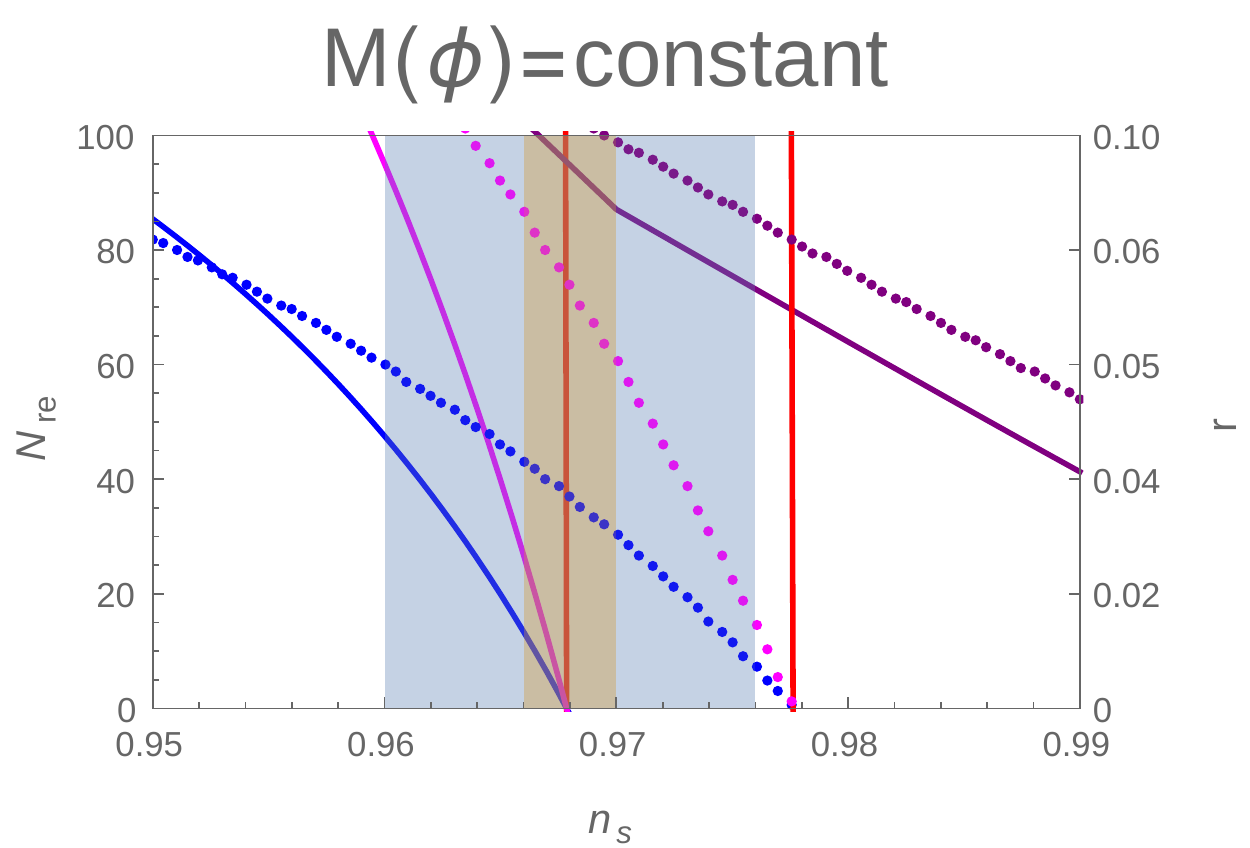}
 		\includegraphics[width=005.0cm,height=03.80cm]{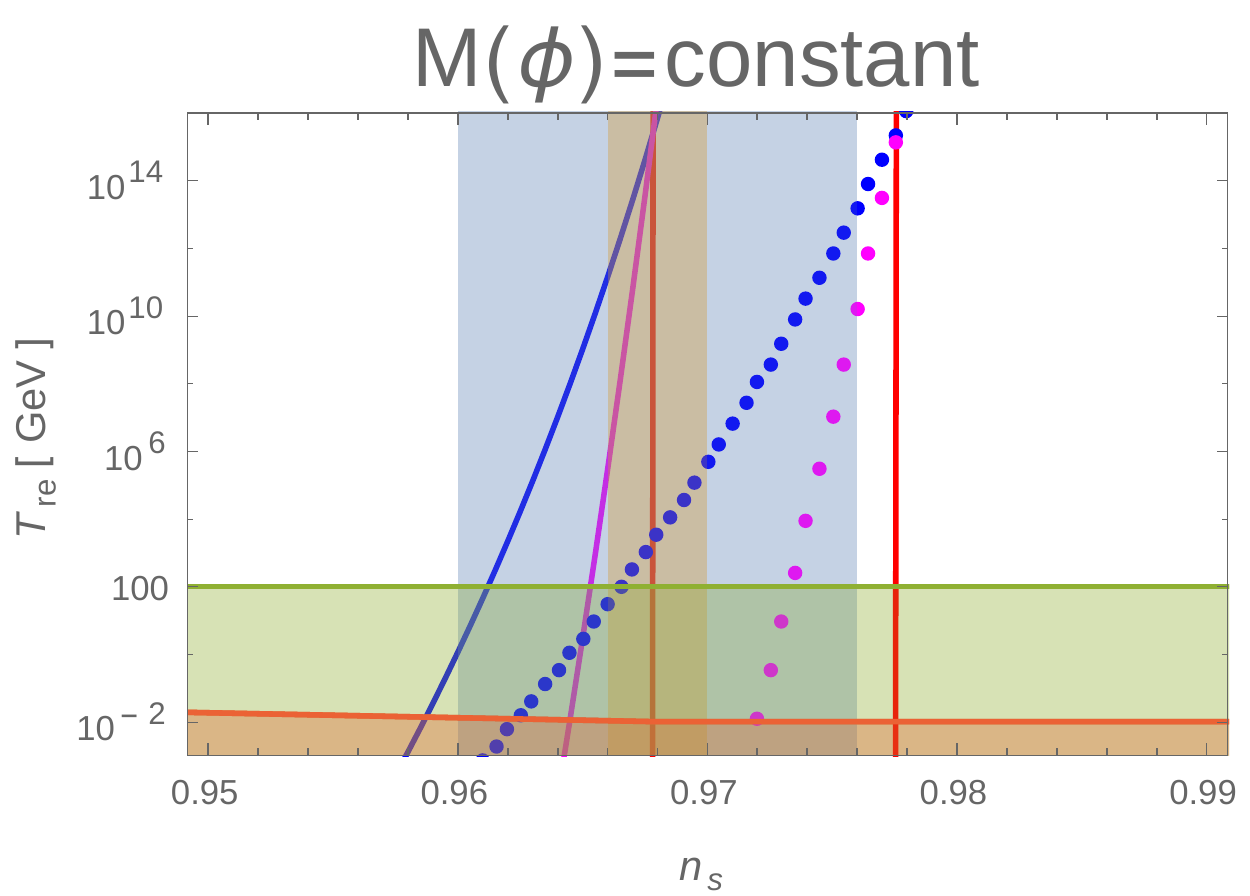}\\
 		\includegraphics[width=005.0cm,height=04.02cm]{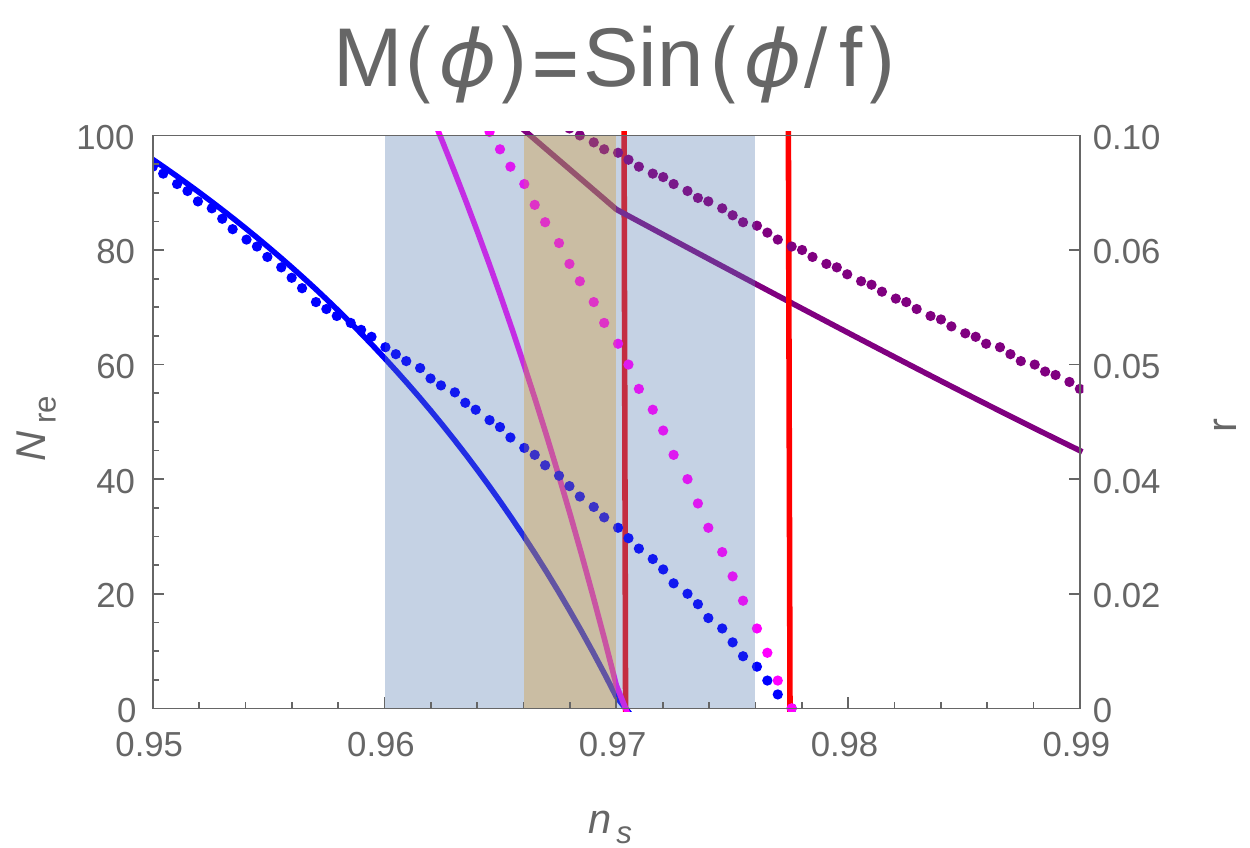}
 		\includegraphics[width=005.0cm,height=03.8cm]{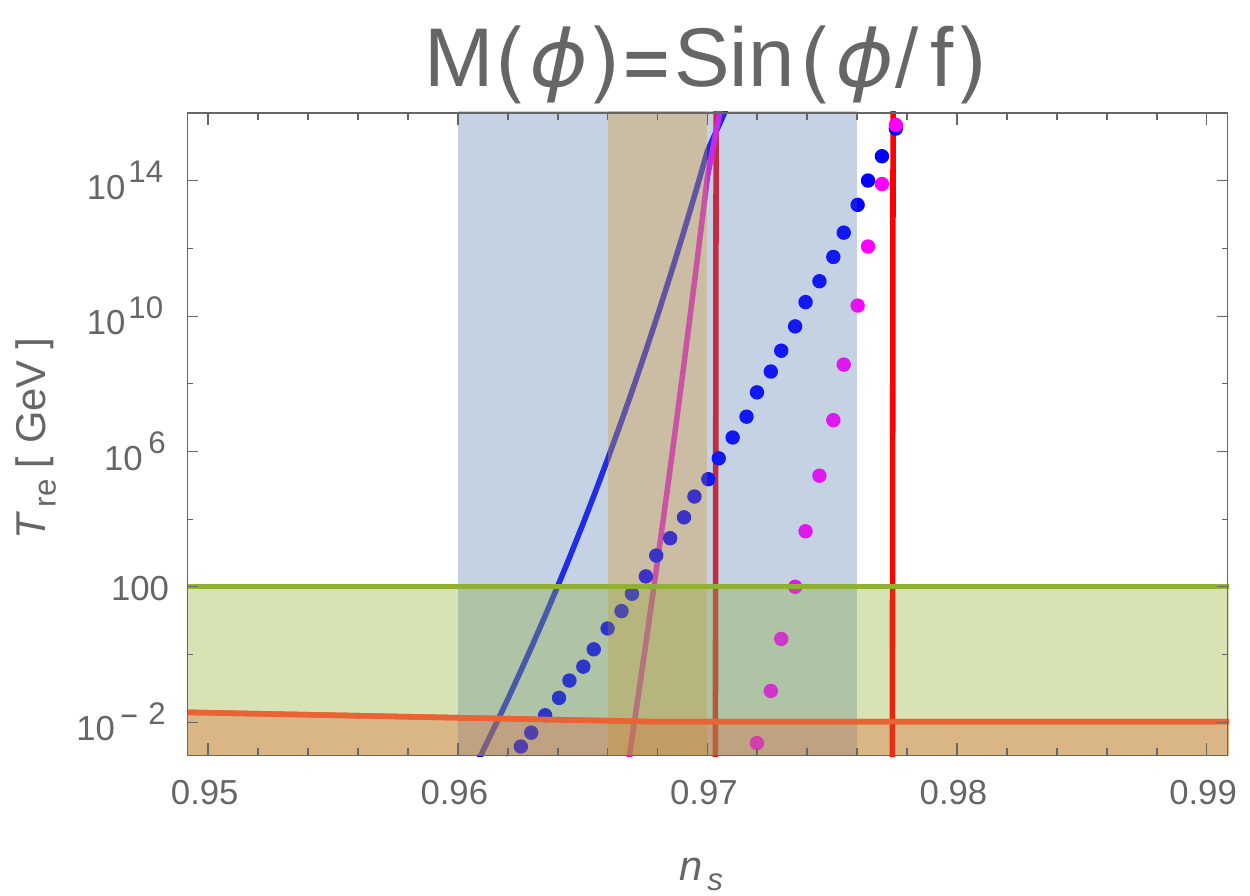}
 		\caption{\scriptsize Variation of $(N_{re}, T_{re})$ as a function of $n_s$ have been plotted.
 		 Blue an red plots are for the single reheating phase after the end of inflation for $\omega_{re} =(0,1/3)$ during reheating. The magenta line corresponds to the two phase 
 			reheating process with the theoretically motivated set of equation of state parameters, 
 			$(\omega^1_{re}=0,\omega^2_{re} = 1/3)$, and equal number of e-folding parameters $N^1_{re}=N^2_{re}$. 
 			The light blue shaded region corresponds to the $1 \sigma$ bounds on $n_s$ from Planck. 
 			The brown shaded region corresponds to the $1 \sigma$ bounds of a further CMB experiment with 
 			sensitivity $\pm 10^{-3}$ \cite{limit1,limit2}, using the same central $n_s$ value as
 			Planck. Temperatures below the horizontal red line is ruled out by BBN. The deep green 
 			shaded region is below the electroweak scale, assumed 100 GeV for reference.} 
 		\label{non-osc}
 	\end{center}
 \end{figure}
 A particular scale $k$ exiting the horizon during inflation and re-entering the horizon during usual cosmological evolution provides us an important 
 relation  $k = a_0 H_0 = a_k H_k$. 
 Where, $(a_{re}, a_0)$ are the cosmological scale factor at the end of the reheating phase and at the present time respectively.  
$(N_k,H_k)$ are the efolding number and the Hubble parameter respectively for the aforementioned scale $k$ 
which exits the horizon during inflation. 
$g_{re}$ is the number of
relativistic degrees of freedom after the end of reheating phase.
$T_0$ is the current value of the CMB temperature.  
For further calculation, we define a quantity, $\gamma = N^2_{re}/N^1_{re}$. We identify
the scale of cosmological importance $k$ as the pivot scale of PLANCK,  $k/a_0 = 0.05 Mpc^{-1}$, and
the corresponding estimated scalar spectral index to be $n_s = 0.9682 \pm 0.0062$.

The background inflationary dynamics of all the models under consideration do not explicitly depend upon axion decay constant $f$ but on an emergent parameter ${\cal A}$. Hence our subsequent discussions on constraints from reheating will remain same for any value of $f$ above critical value $f_c$ while keeping ${\cal A}$ constant. Before we quantify the constraints on our model, general description of our figure are as follows:  Throughout the analysis we will consider two physically realizable values of $\omega_{re} = (0, 1/3)$. It can be shown
that if we consider single phase reheating process, for $\omega_{re}=1/3$, both the quantities, $T_{re}, N_{re}$, become indeterministic. 
This fact corresponds to all the vertical solid red lines in $(n_s~vs~T_{re})$ and $(n_s~vs~N_{re})$ plots. All the dotted or solid blue 
curves correspond to $\omega_{re} = 0$. On the other hand, all the dotted and solid magenta curves correspond to the two stage reheating process
with $(\omega^1_{re}=0,\omega^2_{re} =1/3)$, and $\gamma =1$. It is worth mentioning that by tuning either $\gamma$, or $\omega_{re}$ within $(0,1/3)$ the magenta curve will always remain within the blue curve and the vertical red line. On the same plot of $(n_s~vs~N_{re})$, we also plotted $(n_s~vs~r)$ corresponding to the brown curves. The solid brown curve corresponds to ${\cal A} =100$, for the  sub-Planckian models. For the super-Planckian models, solid brown curve corresponds to the 
bunch of curves which are in the middle of the three bunches. For super-Planckian model, $r$-prediction is almost independent of ${\cal A}$ value. If we consider PLANCK's bound of $r<0.07$, one can easily discard some region of the parameter ${\cal A}$. Each bunch of (red, magenta, blue) curves apparently emanating from a point corresponds to a particular value of ${\cal A}$. Therefore, at least for the super-Planckina models, as we go towards higher value of ${\cal A}$ it will given the overall description of all the plots. We now set to constrain the
model parameters and corresponding prediction for $(T_{re},N_{re})$ for two different kind of models. 

\begin{figure}
	\begin{center}
		\includegraphics[width=005.0cm,height=04.0cm]{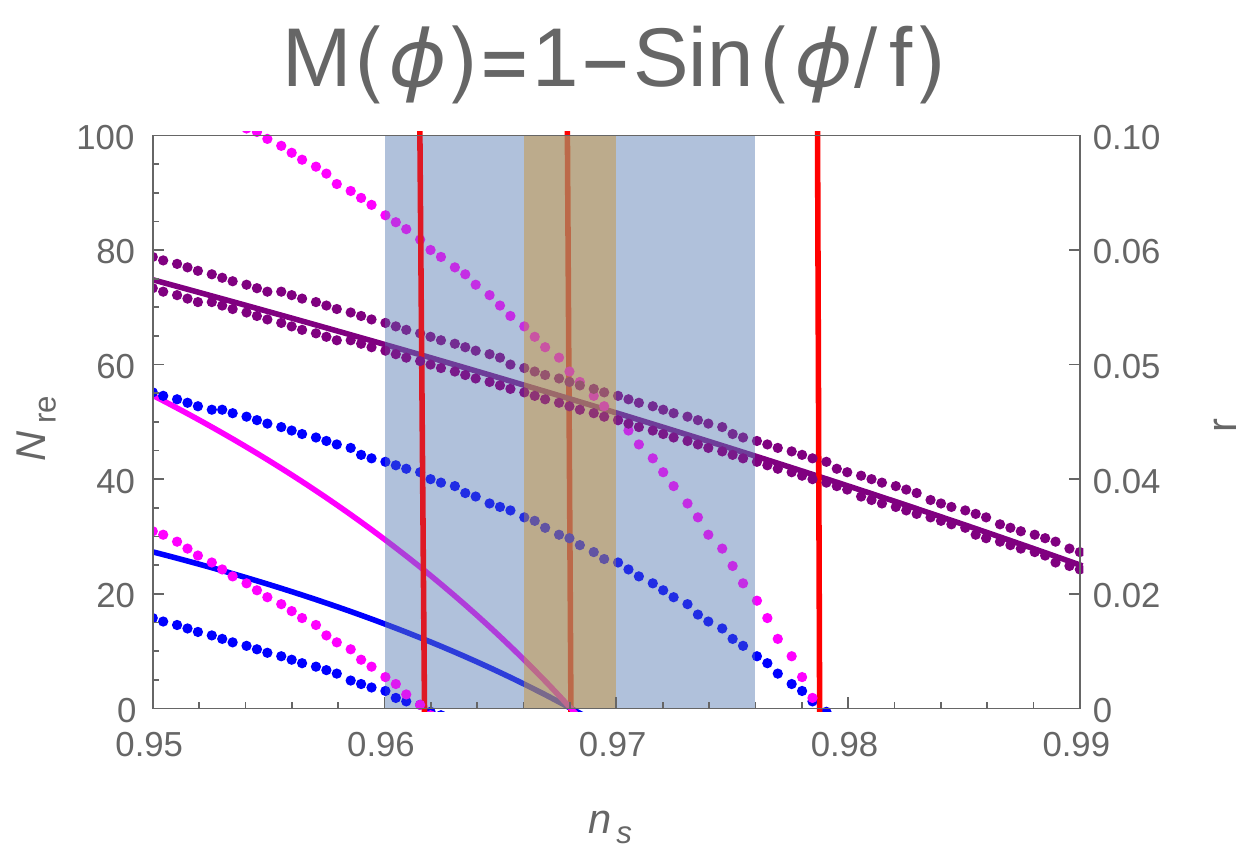}
		\includegraphics[width=005.0cm,height=03.8cm]{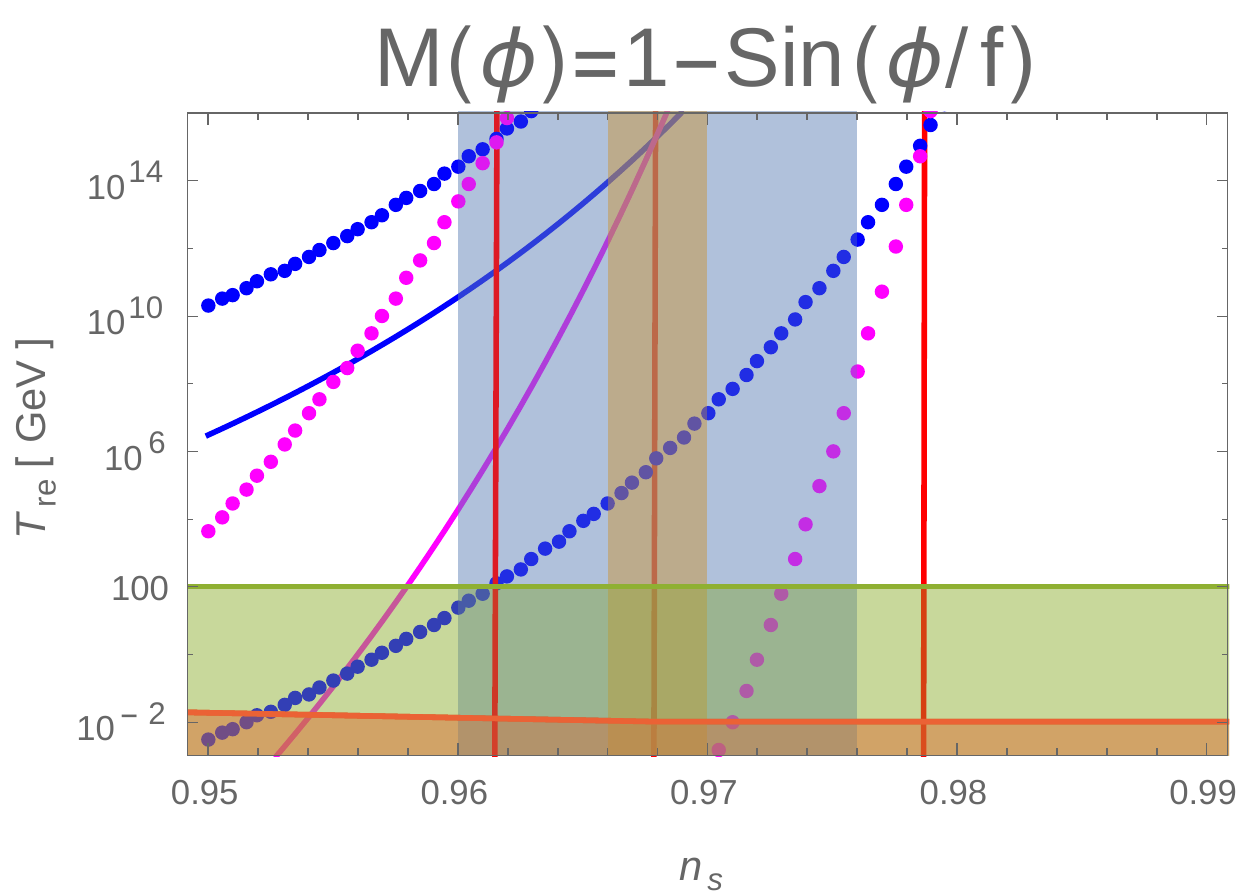}\\
		\includegraphics[width=005.0cm,height=04.02cm]{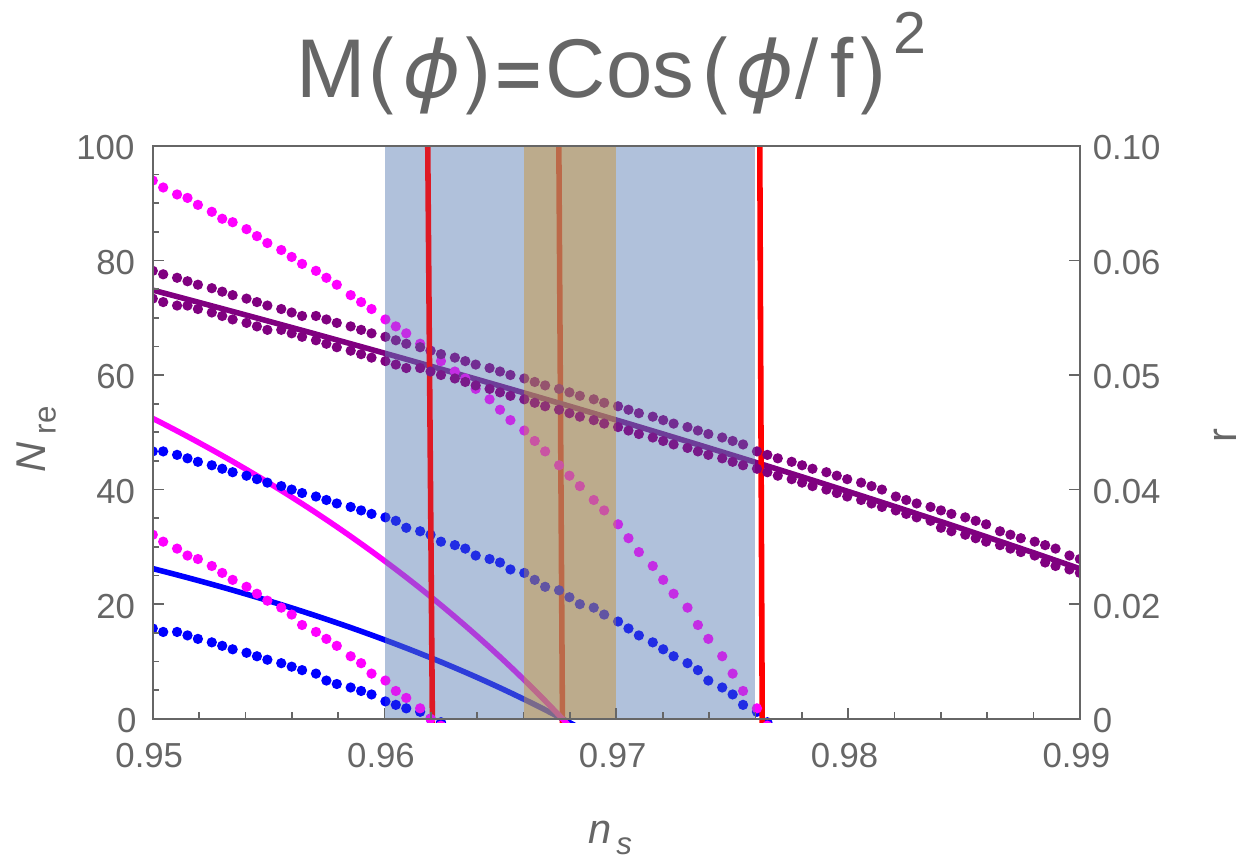}
		\includegraphics[width=005.0cm,height=03.8cm]{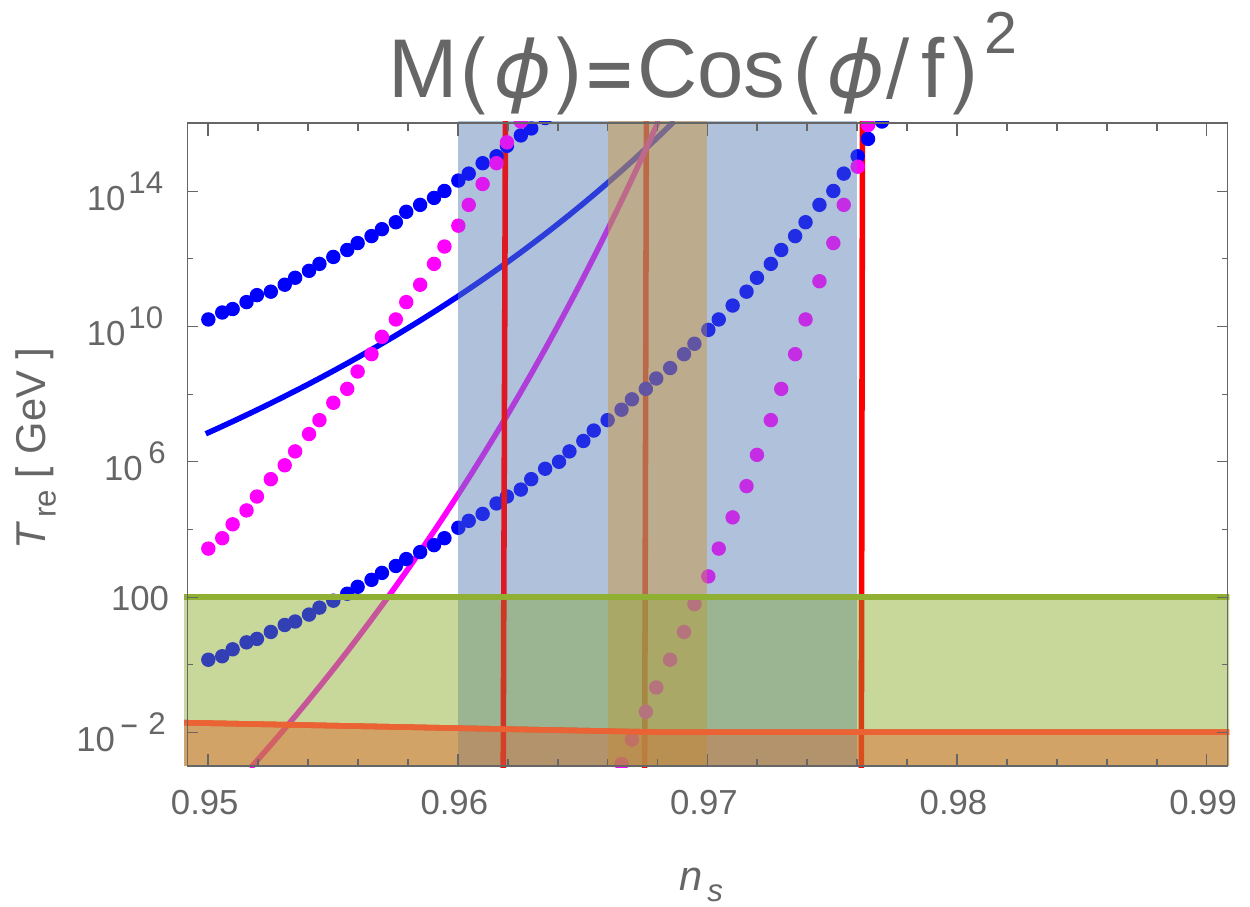}
		\caption{\scriptsize Variation of $(N_{re}, T_{re})$ as a function of $n_s$ have been plotted for two different models. All the other parameters are taken to be same as in the previous plot fig.\ref{non-osc}.} 
		\label{osc}
	\end{center}
\end{figure}

\begin{figure}
	\begin{center}
		\includegraphics[width=005.0cm,height=04.0cm]{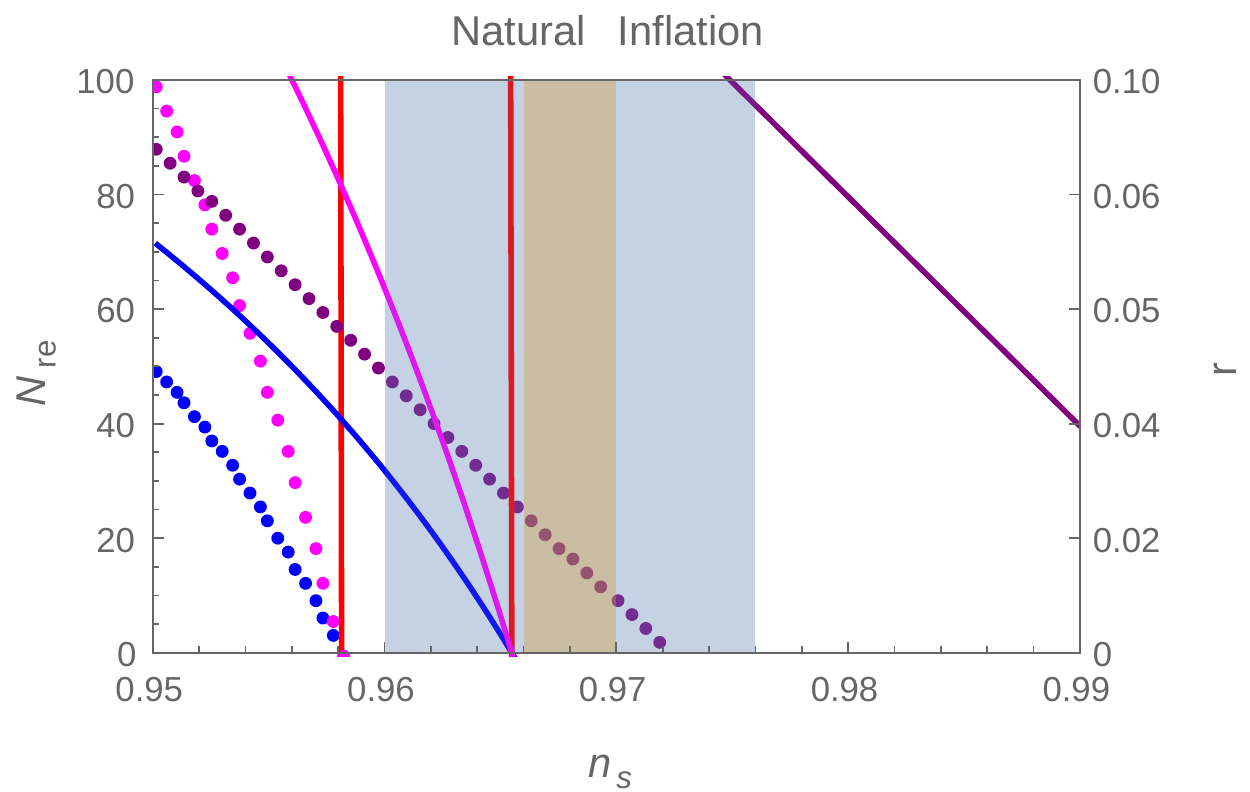}
		\includegraphics[width=005.0cm,height=03.82cm]{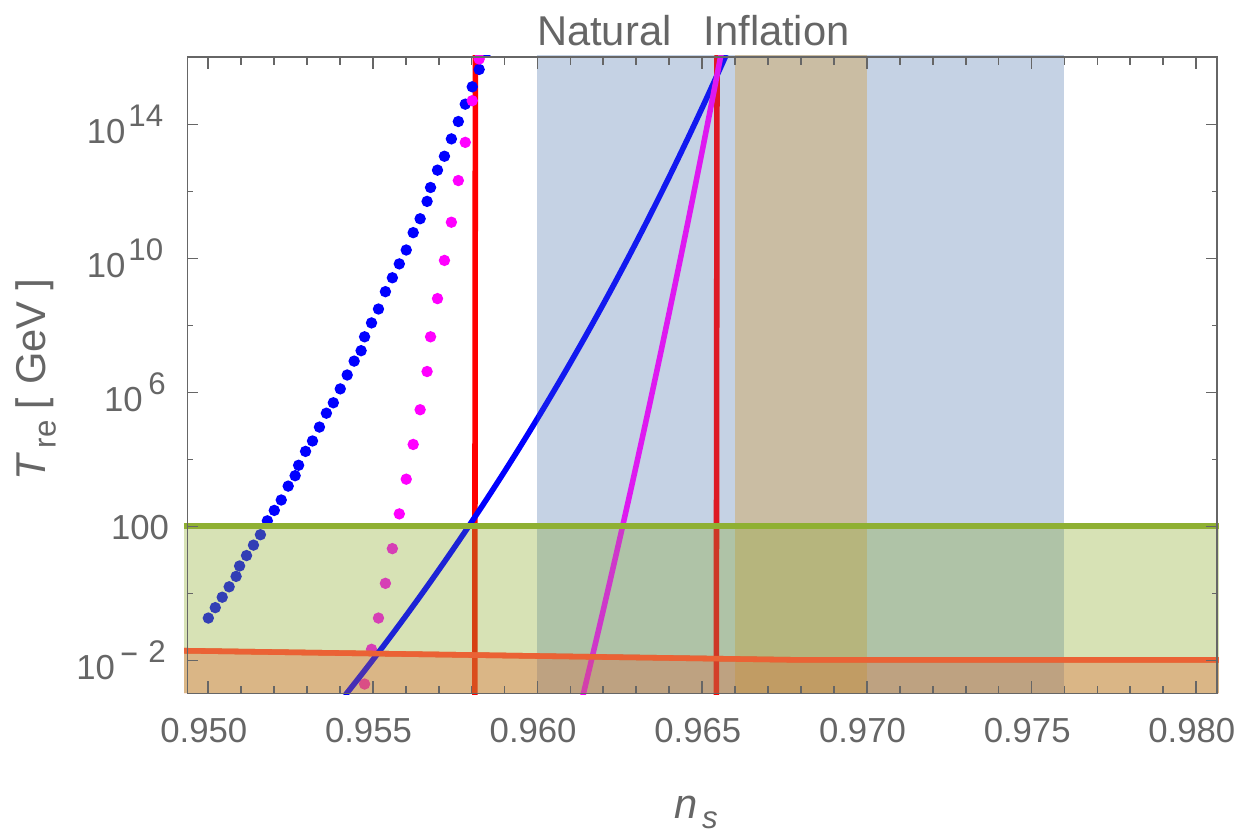}
		\caption{\scriptsize Variation of $(N_{re}, T_{re})$ as a function of $n_s$ for usual axion inflation model without any galileon term.} 
		\label{natural}
	\end{center}
\end{figure}

\subsection{Sub-Planckian models: $M(\phi) =\{\mbox{constant}, \sin(\phi/f),\sin^2(\phi/f)\}$}

As we have described already, sub-Planckian models are those which can explain CMB observation even for sub-Planckian axion decay constant. For those models we can have axion decay constant $f$ as low as  Pecci-Quinn symmetry breaking scale $f=10^9 \sim 10^{12} \mbox{GeV}$. However, from the effective field theory point of view, we must limit the value of $f$ above the inflationary energy scale $\Lambda \simeq 10^{-2}$ in the unit of Planck. As explained before, for these models, the
axion field will not have an oscillating phase after inflation. Therefore, we have to employ the instant preheating
scenario which we have extensively discussed in the next section. Important to mention that for all these $M(\phi)$'s, we will certainly 
have coherent oscillation, but with the axion decay constant $f\gtrsim f_c \simeq {\cal O}(1) M_p$. For illustration, we have plotted only for first two models. As we have found, for both the models the critical value
at which the axion field starts to oscillate after the inflation are $f_c \simeq (3, 1) M_p$ in Planck unit for
${\cal A} = (76,92)$ respectively. For illustration we have considered ${\cal A}=(100,60)$ for $M(\phi) =1$ and ${\cal A} =(100,70)$ for
$M(\phi) = \sin(\phi/f)$ in the fig.\ref{non-osc}. For both the models, ${\cal A} =100$ corresponds to the left bunch of curves emanating from a point on $r=0$ line.
As we further increase the value of ${\cal A}$, the left bunch or in other words the $\omega = 1/3$ line does not move further left.
This essentially means, for a particular value of efolding number N, we have a minimum value of $n_s$ as a function of ${\cal A}$. This can also be seen from the Fig.(\ref{planckplot}). 
   
\subsection{Super-Planckian models: $M(\phi) =\{1-\sin(\phi/f), \cos^2(\phi/f) \} $}

In this section, we discuss those models which fit well with the PLANCK observation for super-Planckian axion decay
constant, Fig.(\ref{planckplot}). 
Let us re-emphasize again that for these models also we have critical values of $f_c\simeq (5, 4) M_p$. However, below this value the inflaton encounters singularity before the inflation ends. Therefore, we only have oscillatory solution for these models which are observationally viable. In order to compare, we have also plotted 
$(N_{re},T_{re})$ with respect to $n_s$ for usual natural inflation as shown in Fig.(\ref{natural}). Therefore, from the $(r ~vs~n_s)$(brown curve) superimposed with the $(N_{re}~vs~n_s)$ curve, clearly disfavored the conventional natural(axion) inflation. Going beyond the natural(axion) inflation is necessary. In any case for illustration we have considered ${\cal A}=(160, 195, 210)$ for $M(\phi)= 1- \sin(\phi/f)$ and ${\cal A}=(130, 150, 160)$ for $M(\phi)= \cos(\phi/f)^2$ as shown in Fig.(\ref{osc}). Where increasing the value of ${\cal A}$ is equivalent to going from the right bunch of curve to left bunch. Therefore, within $1\sigma$, one can clearly discard ${\cal A} >(160,210)$ for $(IV,V)$ models respectively. One can easily see that our generalized natural inflationary models fall within the $1 \sigma$ region in
$(n_s,r)$ space provided by PLANCK as opposed to the natural inflation with usual kinetic term.

\section{Non-oscillating axion: Instant preheating}\label{instant_preheating}
\hspace{0.5cm} Reheating phase after the inflation is a stage when the energy stored in the inflaton field is transferred into the matte field we see today. As we have noted, we get two types of scalar field solution after inflation. For oscillatory solution, the standard mechanism of reheating works well in our model\cite{modnat, massive}. But for the cases when the scalar field does not have coherent oscillation after the inflation, we need to invoke alternative mechanism to reheat our universe. We see that instant preheating, originally proposed as an alternative to preheating mechanism for no-oscillatory inflation models, could be important. This mechanism has been successfully implemented to reheat the universe in quintessential inflation \cite{campos, gr-qc-0307068, sami_sahni, panda_sami}. In the following sections, we study instant preheating for sub-Planckian model for $f< f_c \simeq {\cal O}(1) M_p$. As mentioned before, we consider two types of coupling between the reheating field and the inflaton and compare their results. We will see how the higher derivative coupling term plays the important roll in our analysis.  \\

\subsection{Conventional Shift Symmetry Breaking Coupling}
\begin{figure}[t!]
 	\centering
 	\subfigure[]{\includegraphics[scale=0.35]{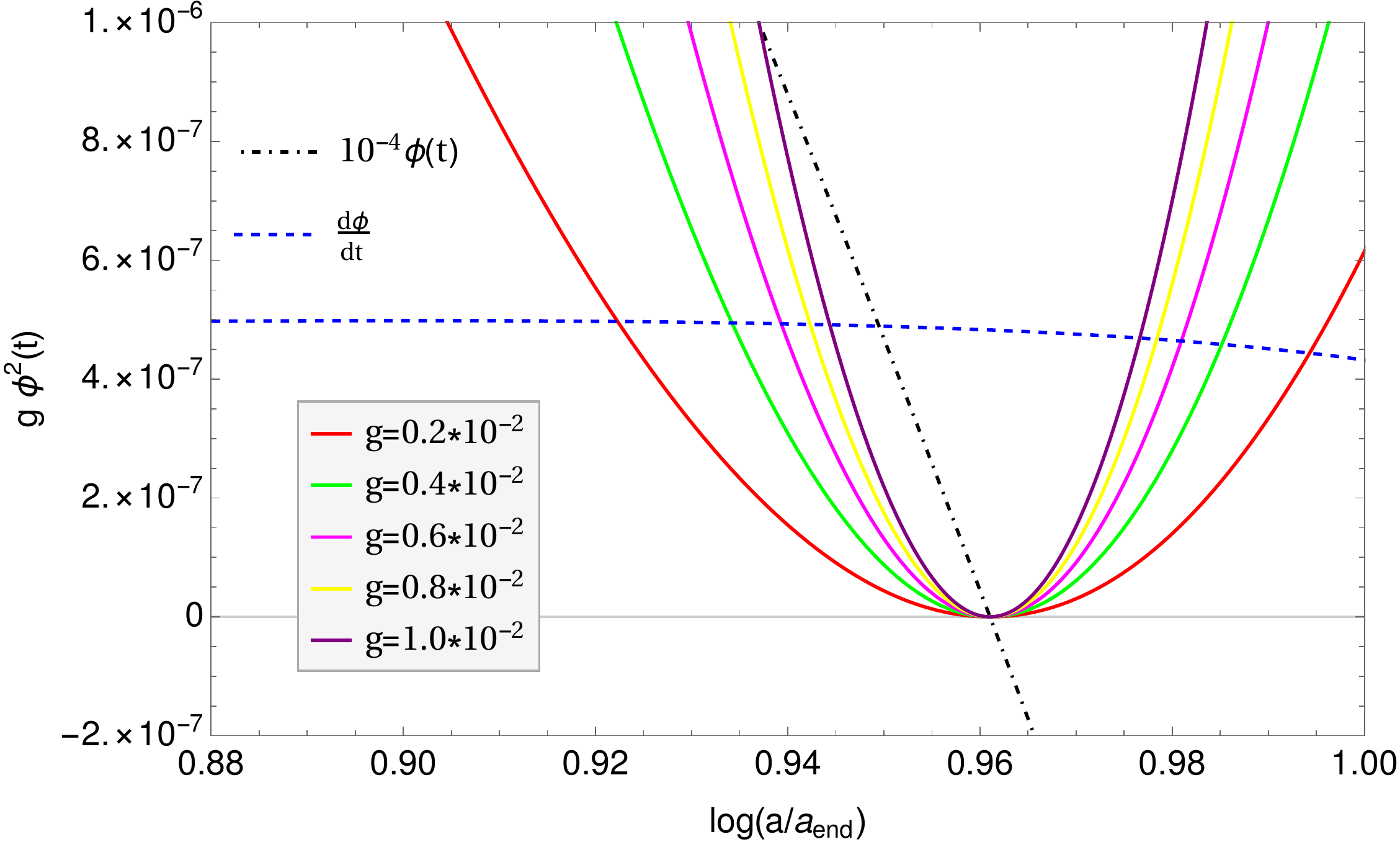}}
 	\subfigure[]{\includegraphics[scale=0.35]{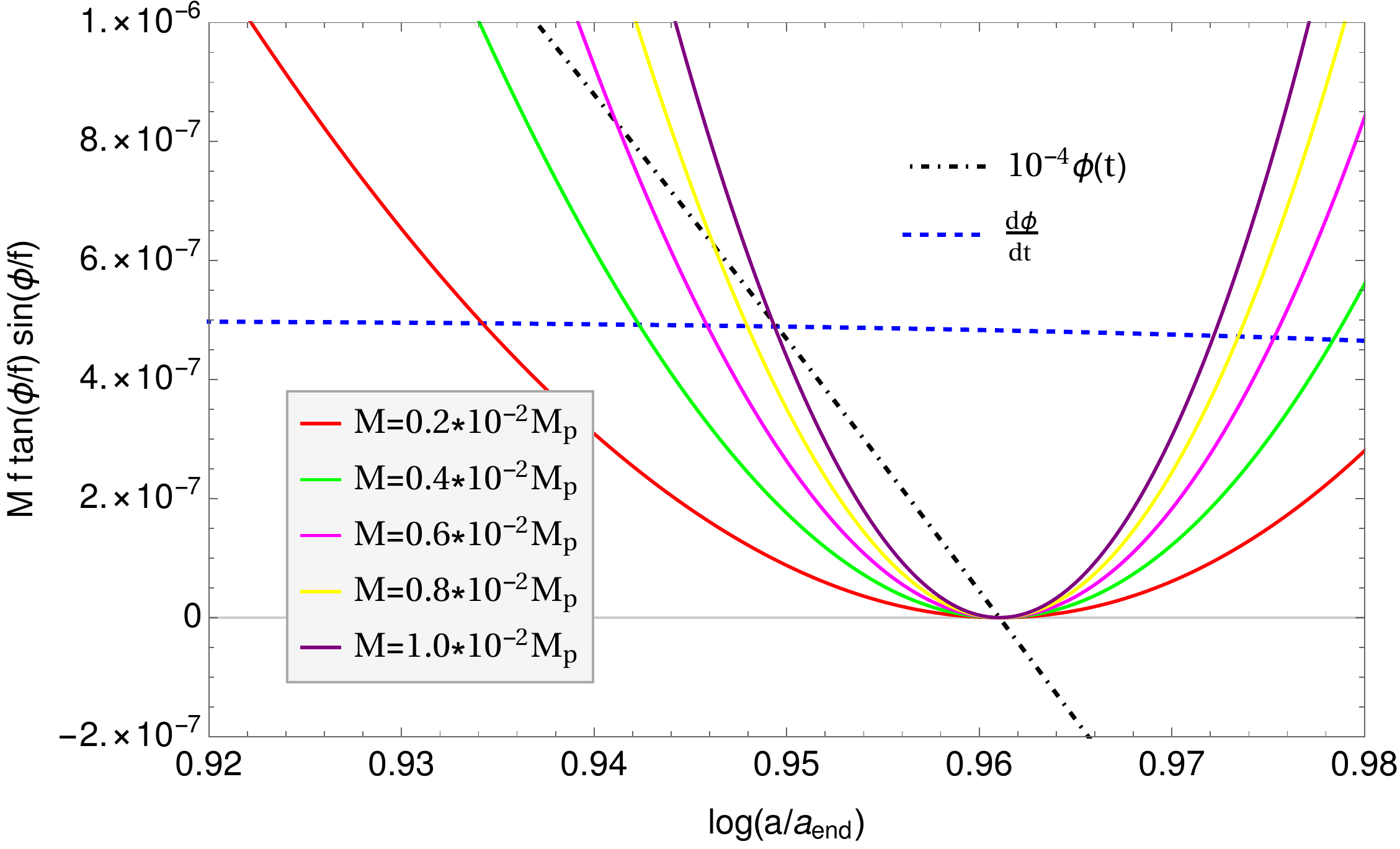}}
 	\caption{\label{rn} \scriptsize Figure shows the region where the adiabaticity condition is violated, which is evidently the region where $\phi(t)$ crosses zero (black dashed line), Fig.(a) is for the coupling $ g^2 \phi^2 \chi^2$, while Fig.(b) is for $M^2sin^2(\phi/f)\chi^2$ }
 	\label{inst-preheat}
 \end{figure}

Let us start by considering the interaction Lagrangian as of the original work of Felder-Kofman-Linde. Where the inflation $\phi$ interacts with another scalar field $\chi$ which decays to a fermion field $\psi$. We write the interaction Lagrangian as
\bea
\mathcal{L}_{int} = -\frac{1}{2}g^2 \phi^2 \chi^2 - h \bar{\psi} \psi \chi
\label{int-L} ,
\eea
where the couplings are supposed to be positive with $g,h < 1$ in order for the perturbation treatment to be valid. It is evident that the above Lagrangian does not respect the shift symmetry of our original Lagrangian. We will consider a special shift symmetric case in the next subsection.
 For simplicity, we will consider the $\chi$ particles do not have a bare mass,
while its effective mass is provided by the inflation field as
 \bea
m_{\chi}(\phi) = g |\phi| .
\label{mchi}
\eea
The production of the $\chi$ initiate as $m_{\chi}$ starts changing non-adiabatically after the end of inflation.
\bea
|\dot{m_{\chi}}| \gtrsim m_{\chi}^2  ~~ or, |\dot{\phi}|\gtrsim g\phi^2
\eea
In the Fig.\ref{inst-preheat}-a, we see the region where the adiabatic condition is violated for different values of dimensionless coupling parameter $g$. Above condition implies that
\bea
|\phi| \lesssim |\phi_{prod}| = \left(\frac{\dot{\phi}_{end}}{g}\right)^{\frac{1}{2}}
\eea
Now, to estimate $\dot{\phi}$ analytically, we assume the slow-roll condition to hold till the end of the inflation.
This is where the higher derivative term in our model plays the role. Using eq.[\ref{phidotend}], we find that
\bea
\begin{split}
| \dot{\phi}_{end}| \simeq \left( \frac{2 \epsilon_{end} M_p}{3 \sqrt{3}  M(\phi)} \right)^{\frac{1}{3}}(V_{end})^{\frac{1}{6}} &= \left( \frac{2 M_p}{3 \sqrt{3}  M(\phi)} \right)^{\frac{1}{3}}(V_{end})^{\frac{1}{6}}\ &= 
P(\phi)(V_{end})^{\frac{1}{6}} ,
\end{split}
\label{phiend}
\eea
where, we have denoted
\begin{equation*}
P(\phi) = \left( \frac{2 M_p}{3 \sqrt{3}  M(\phi)}  \right)^{\frac{1}{3}}
\label{Pphi}
\end{equation*}
It would more intuitive to express the above expression of $P(\phi)$ in terms of the derived parameter, $\mathcal{A}$ that we have defined earlier. This reads as
\begin{equation}
	P(\phi) = 0.72 \frac{1}{\mathcal{A}^{\frac{2}{3}}} \frac{f}{M_p} \left[\frac{\Lambda^{4}}{\tilde{M}(\tilde{\phi)}}\right]^{\frac{1}{3}}
	\label{PphiA}
\end{equation}
The production time for $\chi$ particles can be estimated as
\bea
\Delta t_{prod} \sim \frac{|\phi_{prod}|}{|\dot{\phi_0}|} \sim \frac{1}{\sqrt{g P(\phi) (V_{end})^{\frac{1}{6}}}}
 \eea
where $|\dot{\phi_0}|$ is the velocity of the field near the minimum of the effective potential. Let us now make an estimate of the numerical values of the quantities involved. We can use the relation (\ref{PR}) and the observed value of the scalar power spectrum to determine the value of the inflaton potential at the end of the inflation. Taking a sample value of the KGB scale $s \sim 10^{-5} M_p$, one finds  $V_{end} \sim 6 \times 10^{-6} M_p^4$. In order for the particle production to occurs within a very short period of time, we must satisfy $\Delta t_{prod}/s <1$ which provides lower bound on $g$ as $g > 10^{-6}f_b$. 
Where $f_b$ is the axion decay constant in unit of $M_p$. Using these values, we find the value of $\dot{\phi}_{end} \sim 10^{-6}M_p^2$. Hence to get $\phi_{prod} < M_p$, we must choose $g > 10^{-6}$.
Using uncertainty relation one can estimate the momentum $k_{prod} \simeq (\Delta t_{prod})^{-1} \sim (g P_{end})^{1/2} V_{end}^{1/12}$ of $\chi$ particles which will be be created non-adiabatically, and \cite{inst, kofman}, the occupation number of $\chi$ particles jumps from zero to
\bea
n_{k} \simeq exp(-\pi k^2/k_{prod}^2 )
\eea
during the time interval $\Delta t_{prod}$. The number density can be estimated to be
\bea
n_{\chi} = \frac{1}{2 \pi^2} \int_{0}^{\infty} k^2 n_{k}dk \simeq \frac{k_{prod}^3}{8 \pi^3} \simeq \frac{(g P_{end})^{\frac{3}{2}} V_{end}^{\frac{1}{4}}}{8 \pi^3}
\eea
and the energy density of the $\chi$ particles can be found to be
\bea
\rho_{\chi} = m_{\chi}n_{\chi} \left( \frac{a_{end}}{a} \right)^3 = \frac{(g P_{end} V^{\frac{1}{6}})^{\frac{3}{2}}}{8 \pi^3} g |\phi| \left( \frac{a_{end}}{a} \right)^3
\label{rhochi}
\eea
where the $(a_{end}/a)^3$ term corresponds to the dilution of the energy density due to cosmic expansion.

In numerical calculation, we found that the value of $g$ as small as $\mathcal{O}(10^{-3})$ satisfy the adiabaticity condition mentioned above. And with this value of coupling constant, we can readily 
check that production of the $\chi$ particle is indeed instantaneous. Now, if the quanta of the $\chi$-field were to thermalized into radiation instantly, the radiation energy density would become
\bea
\rho_{r} \simeq \rho_{\chi} \sim \frac{(g P_{end}) V^{\frac{1}{6}})^{\frac{3}{2}}}{8 \pi^3} g \phi_{prod} \sim \frac{g^2 P^2(\phi) V_{end}^{\frac{1}{3}}}{8 \pi^3} . 
\eea 
Now the efficiency of instant preheating can be parametrized by the following ratio,
\begin{eqnarray}
\nno
 \frac{\rho_r}{\rho_{\phi}} &\simeq& 2\times 10^{-3} \left(\frac{g}{\mathcal{A}^{\frac{2}{3}}} \right)^2 \left( \frac{f}{M_p}\right)^2 \left( \frac{1}{\tilde{M}_{end}}\right)^{\frac{2}{3}}  \simeq 5\times 10^{-6} g^2 .
 \label{ratio_sbc}
\end{eqnarray}
Where, in the final numerical value we considered order of magnitude values of all the parameters ${\cal A}\simeq 100, f \simeq 0.1 M_p$, and $\tilde{M}_{end}\simeq 1$.  
To this end let us quote the above ratio of energy densities for the Quintecence inflation model where also one does not have oscillatory phase after the inflation \cite{sami_sahni} 
\begin{equation}
\frac{\rho_r}{\rho_{\phi}} \simeq 10^{-2} g^2 .
\end{equation} 
This result can be proved to be generically true for conventional canonical inflation model with the aforementioned interaction eq.\ref{int-L} term. Therefore, comparing with the usual model, we conclude that for G-axion inflation model instant preheating will not be efficient enough as it is suppressed by the same emergent parameter ${\cal A}$ which played the important role in obtaining the sub-Planckian axion decay constant $f$. In addition, we also have dimensionless coupling constant $g$ which is constrained to be small to avoid strong coupling problem. 

 So far we have not mentioned anything about the fermionic coupling parameter $h$. Even though the particle production at the instant of first zero crossing of inflaton, does not seem efficient, for completeness let us constraint the possible value of $h$ for the instant preheating mechanism to work. The decay rate reheating field into radiation is given by 
$\Gamma_{\psi \bar{\psi}} = h^2 m_{\chi}/8\pi$, with $m_{\chi}= g|\phi|$. Now for instant preheating to work it must be larger than the expansion rate of the universe at instant of preheating 
\bea
\Gamma_{\psi \bar{\psi}} \gtrsim H_{prod} \implies h^2 \gtrsim 8 \pi \frac{H_{prod}}{g \phi_{prod}}
\eea
The order of magnitude of $h$ can be estimated for some sample value of $g$ we can have
\begin{eqnarray*}
 h \gtrsim 0.3\hspace{0.5cm} \text{for}\hspace{0.5cm} g \sim 10^{-2}\\
 h \gtrsim 0.9\hspace{0.5cm} \text{for}\hspace{0.5cm} g \sim 10^{-3}
\end{eqnarray*}

As we have concluded from our analysis, instant preheating mechanism turns out to be inefficient for the conventional shift symmetry breaking coupling considered above. We want to see if any other type such as shift symmetric coupling can give some enhancement in the efficiency. 

\subsection{Shift Symmetric Coupling}
As we have started with shift symmetric theory, for completeness, we express the main result for the following shift symmetric coupling
 \be
 \mathcal{L}_{int} = -\frac{1}{2} M^2~ sin^2\left(\frac{\phi}{f}\right)\chi^2 - h \bar{\psi}\psi . \chi
 \ee
 Where $M$ dimensionful coupling parameter. In the fig.\ref{inst-preheat}-b, we plotted the region where the adiabaticity is lost.  Follow the same methodology discussed before and under some reasonable assumption and approximation, we arrived at the following ratio of produced energy density over the inflaton energy density, 
 \begin{eqnarray}
 	\frac{\rho_r}{\rho_{\phi}} \simeq 2 \times 10^{-3} \left(\frac{M}{M_p\mathcal{A}^2}\right) \left(\frac{\Lambda}{M_p} \right)^2 \left( \frac{1}{\tilde{M}_{end}} \right) \simeq 2 \times 10^{-7} M ,
 	\label{ratio_ssc}
 \end{eqnarray}
Where again for final numerical value, we have chosen ${\cal A}=100, \Lambda =10^{-2} M_p$, and $\tilde{M}_{end}\simeq 1$. Therefore, we again see the suppression by the factor ${\cal A}$. 

Our observation in this section is that the instant preheating mechanism seemed to be inefficient in transferring the energy from inflaton to reheating field for higher derivative driven G-axion inflation. 
This negative result can be attributed to the fact that the velocity of the inflaton field near the end of inflation is suppressed by the parameter ${\cal A}$ as opposed to the usual axion inflation scenario. Therefore, the adiabaticity violation turned out to be very weak near the zero crossing of the inflaton field. Hence, the particle production became inefficient compared to the usual canonical inflation models.

\section{Conclusions}

In this paper we have studied the cosmology of a modified axion inflation model with a specific form of higher derivative kinetic term. We call it G-axion. As we have already discussed, the usual axion inflation is disfavored for its prediction of larger tensor to scalar ratio within $1\sigma$ range of $n_s$ of PLANCK. From our two component reheating constraint analysis shown in
 fig.(\ref{natural}), we also noticed that the canonical axion inflation model is outside the PLANCK limit. Therefore, modification of axion inflation is essential. In this paper we extended our previous analysis \cite{debuaxion,modnat} taking into account some simple choices of Galileon interaction term $M(\phi)X\Box\phi$ in understanding more on the sub-Planckian dynamics.  
One of the important motivations of studying such a modification, of course, is to explain the current observational data by PLANCK. However, another reason is to make the model consistent in the framework of effective field theory.  
With such a higher derivative modification, our model turned out to predict inflationary parameter $(n_s,r)$ within the range of Planck data for sub-Planckian values of all its parameters. However, we have found for some specific choices of functions, $M(\phi)=\{ constant,\sin(\phi/f),\sin^2(\phi/f ) \}$, the sub-planckian axion decay constant yields singular behavior in the scalar field dynamics specifically after the inflation. We dubbed them as non-oscillatory G-axion model. The existence of such kind of singular behavior in cosmological context has been discussed in recent
 studies\cite{Bains:2015gpv,Easson:2016klq}. The singularity behavior emerges from the kinetic function $\Delta(t)$ when approaching towards zero. The significance of the singularity in Kinetic Gravity Braiding theories has been discussed.  Exploring this property for further analysis in the context of dark energy will be interesting, and we left it for our future studies. The issues of stability and superluminal propagation speed has been briefly discussed. It has been found that there exits parameter space when such instabilities will arise. The possible resolution for this issues in the light of \cite{Cai:2016thi,Cai:2017dyi,Easson:2018qgr} could be very interesting for our future work. Nonetheless, it has been found that these sub-Planckian models provide successful inflation and right after crossing
 the zero in the field space it hits the singularity. Since these models give successful inflation, we employed the instant preheating mechanism to reheat our universe. However, it turned out that the mechanism under study is not efficient enough. The amount of energy transferred from inflaton to the reheating field is suppressed by the same factor ${\cal A}$ which helped us to make the model consistent with the PLANCK observation on inflationary observables. We have studied two different types of coupling namely, shift symmetric and conventional broken shift symmetric, to study preheating. For both types of coupling energy transfer turned out the be inefficient. Therefore, we need further study to understand the sub-Planckian G-axion model.  
 Interestingly, with regards to the latest observation made by PLANCK, we found out super-Planckain G-axion models for  $M(\phi)=\{\cos^2(\phi), 1-\sin(\phi)\}$ which are very well fitted as opposed to the conventional axion inflation. For those two models we have prediction of $r\simeq 0.03 \sim 0.05$ within the $1\sigma $ range of $n_s$ which could be detectable in the near future CMB experiments. For super-Planckian G-axion models, the inflaton oscillates after the end of inflation. We, therefore, have studied model independent reheating constraint analysis. 
 
 A particularly interesting point we would like to mention is the role played by the axion decay constant $f$ in the dynamics of G-axion inflation. Depending on the value of $f$, dynamics could be either usual kinetic term dominated or the galileon term dominated. Therefore, an interesting regime of $f$ exists when both terms play the role. In terms of the emergent parameter $\mathcal{A}$, the condition for inflation dominated by KGB term translated into the following approximate relation: $f \ll \mathcal{A}^{1/2}$. Although this bound on $f$ has a certain degree of dependence on the choice of $M(\phi)$. If the value of the axion decay constant violates the above bound, depending on the initial condition, our numerical computation shows that the inflaton dynamics is dominated by the higher derivative term at the initial stage and subsequently it becomes canonical kinetic term dominated around the end of inflation. This fact of two phases of inflation will certainly have interesting consequences on the observable quantities and CMB spectrum. For instance, in some models of inflation\cite{Lello:2013mfa, Liu:2013iha, Lello:2013awa} a phase of super inflation is introduced at the initial stage of inflation to explain the power suppression in the large angular scales of the CMB. This interesting effect can be naturally explained in our model. We left it for our future study. 
 
 \section{Acknowledgment}
 
 We would like to thank Alexander Vikman for useful comment on the draft. We thank the anonymous referee for useful comments which helps us to improve the work.  We are also thankful to our HEP and GRAVITY group members for numerous vibrant discussions.

\end{document}